\documentclass[fleqn,usenatbib,useAMS]{mnras}

\usepackage{graphicx}	
\usepackage{amsmath}	
\usepackage{amssymb}	
\usepackage{multicol}        
\usepackage{bm}		
\usepackage{pdflscape}	

\newcommand{\km}{\rm\thinspace km}

\newcommand{\cm}{\rm\thinspace cm}

\newcommand{\s}{\rm\thinspace s}
\newcommand{\ks}{\rm\thinspace ks}

\newcommand{\Msun}{\hbox{$\rm\thinspace M_{\odot}$}}

\newcommand{\keV}{\rm\thinspace keV}

\newcommand{\kmps}{\hbox{$\km\s^{-1}\,$}}

\newcommand{\psqcm}{\hbox{$\cm^{-2}\,$}}

\newcommand{\pcmcu}{\hbox{$\cm^{-3}\,$}}

\newcommand{\rg}{\rm\thinspace $r_\mathrm{g}$}

\usepackage[T1]{fontenc}
\usepackage{ae,aecompl}

\usepackage{newtxtext,newtxmath}

\title[X-ray flares from I\,Zw\,1]{Acceleration and cooling of the corona during X-ray flares from the Seyfert galaxy I\,Zw\,1}

\author[D. R. Wilkins \textit{et al.}]{D. R. Wilkins$^{1}$\thanks{Contact e-mail: \href{mailto:dan.wilkins@stanford.edi}{dan.wilkins@stanford.edu}}, 
L. C. Gallo$^{2}$, E. Costantini$^{3,4}$, W. N. Brandt$^{5,6,7}$ and R. D. Blandford$^{1}$
\\
$^{1}$Kavli Institute for Particle Astrophysics and Cosmology, Stanford University, 452 Lomita Mall, Stanford, CA 94305, USA\\
$^{2}$Department of Astronomy and Physics, Saint Mary’s University, Halifax, NS. B3H 3C3, Canada\\
$^{3}$SRON, Netherlands Institute for Space Research, Sorbonnelaan 2, 3584 CA Utrecht, The Netherlands\\
$^{4}$Anton Pannekoeck Institute for Astronomy, University of Amsterdam, Science Park 904, 1098 XH Amsterdam, The Netherlands\\
$^{5}$Department of Astronomy and Astrophysics, 525 Davey Lab, The Pennsylvania State University, University Park, PA 16802, USA\\
$^{6}$Institute for Gravitation and the Cosmos, The Pennsylvania State University, University Park, PA 16802, USA\\
$^{7}$Department of Physics, 104 Davey Lab, The Pennsylvania State University, University Park, PA 16802, USA
}

\date{Accepted 2022 February 11. Received 2022 February 10; in original form 2021 September 29}

\pubyear{2022}

\begin{document}
\label{firstpage}
\pagerange{\pageref{firstpage}--\pageref{lastpage}}
\maketitle

\begin{abstract}
We report on X-ray flares that were observed from the active galactic nucleus I\,Zwicky\,1 (I\,Zw\,1) in 2020 January by the \textit{NuSTAR} and \textit{XMM-Newton} observatories. The X-ray spectrum is well-described by a model comprised of the continuum emission from the corona and its reflection from the accretion disc around a rapidly spinning ($a > 0.94$) black hole. In order to model the broadband spectrum, it is necessary to account for the variation in ionisation across the disc. Analysis of the X-ray spectrum in time periods before, during and after the flares reveals the underlying changes to the corona associated with the flaring. During the flares, the reflection fraction drops significantly, consistent with the acceleration of the corona away from the accretion disc. We find the first evidence that during the X-ray flares, the temperature drops from $140_{-20}^{+100}$\keV\ before to $45_{-9}^{+40}$\keV\ during the flares. The profile of the iron K line reveals the emissivity profile of the accretion disc, showing it to be illuminated by a compact corona extending no more than $7_{-2}^{+4}$\rg\ over the disc before the flares, but with tentative evidence that the corona expands as it is accelerated during the flares. Once the flares subsided, the corona had collapsed to a radius of $6_{-2}^{+2}$\rg. The rapid timescale of the flares suggests that they arise within the black-hole magnetosphere rather than in the accretion disc, and the variation of the corona is consistent with the continuum arising from the Comptonisation of seed photons from the disc.
\end{abstract}




\section{Introduction}
The accretion of matter onto supermassive black holes powers some of the most luminous objects we observe in the Universe; active galactic nuclei, or AGN. A significant fraction of their energy output arises in the form of the X-ray continuum, produced by a compact corona of energetic particles associated with the black-hole magnetosphere and the inner parts of the accretion disc \citep{galeev+79,haardt+91}.

It is hypothesised that the X-ray continuum is produced in the corona by the Comptonisation of ultraviolet seed photons emitted thermally from the accretion disc, producing the observed power law form of the continuum spectrum, cut off exponentially at an energy corresponding to the characteristic temperature of the corona. Much remains unknown, however, about the exact nature of the corona, its precise location and structure, and the physical processes by which it is energised from the accreting material, or even the spin of the black hole. Another outstanding question is the relationship between the X-ray emitting corona and the powerful jets that emanate from radio-loud AGN. Recent studies have shown that in many radio-loud AGN, the X-ray emission still appears to come from a corona analogous to that found in radio-quiet AGN, rather than from the jet itself \citep[\textit{e.g.}][]{zhu+2020,zhu+2021}, but in radio-quiet AGN, there may be a component of the corona that can be associated with a failed jet \citep{propagating_lag_paper, yuan_fluxtubes_2}.

The X-ray emission from the corona is highly variable. Short-timescale variability is frequently observed, with the count rate changing by factors of two or three on timescales of just hours. In addition, many AGN have been seen to transition from higher to lower flux `states' or `epochs' corresponding to changes in the structure of the corona \citep[\textit{e.g.}][]{1h0707_jan11,mrk335_corona_paper}. On top of this, transient phenomena, including bright X-ray flares, are observed. A particularly bright X-ray flare was observed from the AGN Markarian 335 in 2014, during which the corona was found to have been ejected and accelerated away from the accretion disc, akin to a failed jet-launching event \citep{mrk335_flare_paper}.

The X-ray emission from the corona illuminates the inner regions of the accretion disc. In addition to seeing the direct continuum emission from the corona, we therefore see the characteristic `reflection' spectrum, produced when the X-ray continuum is reprocessed by the plasma in the accretion disc \citep{ross_fabian}. The reflection spectrum contains emission lines produced by the accretion disc, notably the iron K$\alpha$ fluorescence line around 6.4\keV, a soft excess of emission below approximately 1\keV, and the `Compton hump' that is observed to peak around 20\keV. The emission lines from the inner accretion disc are broadened by the combination of Doppler shifts (from their orbital motion) and gravitational redshifts (due to the proximity of the emission to the black hole), producing a blueshifted peak and characteristic redshifted wing \citep{fabian+89}.  The shape of the redshifted wing encodes a wealth of information including the spin of the black hole \citep{brenneman_reynolds}, the location and geometry of the corona and the structure of the inner disc \citep{1h0707_emis_paper,understanding_emis_paper}.

A further dimension is added to this picture by the measurement of reverberation time lags in the X-ray emission. Variability of the soft X-ray excess and iron K line that are reflected from the disc is found to lag behind correlated variations in the primary X-ray continuum, owing to the additional light travel time from the corona to the inner accretion disc \citep{fabian+09,reverb_review}. The time lags are short relative to the time for light to cross event horizon scales, indicating that the corona is compact and that the reflected X-ray emission is probing the innermost region around the black hole \citep{demarco+2012,kara+2016}. When coupled with measurements of the broad iron K line, the measured reverberation timescale provides additional constraints on the location and geometry of the corona, notably its scale height above the accretion disc \citep{lag_spectra_paper,cackett_ngc4151,propagating_lag_paper}. 

I Zwicky 1 (I\,Zw\,1) is classified as a narrow-line Seyfert 1 (NLS1) galaxy \citep{gallo_nls1_2018} and is found at redshift $z=0.06$. NLS1 galaxies are AGN characterised by comparatively high mass accretion rates onto relatively less massive supermassive black holes between $10^6$ and $10^{7}$\Msun\ \citep{boller+96}. The mass of the black hole in I\,Zw\,1 is estimated to be between $8 \times 10^6$ and $3\times 10^7$\Msun\ via the width of the H$\beta$ line and optical reverberation mapping \citep{vestergaard+06,huang+2019}. I\,Zw\,1 is classed as a `complex' NLS1 galaxy, weaker in its X-ray emission compared to its optical and UV emission, relative to the `simple' NLS1 galaxies, but exhibiting extreme variability, showing sharp peaks and drops in its X-ray light curve \citep{gallo_nls1}.

Detailed measurements of X-ray reverberation from the inner accretion disc in I\,Zw\,1 reveal a corona that is composed of two components \citep{gallo_1zw1_1,gallo_1zw1_2,1zw1_corona_paper}. Part of the corona extends over the inner accretion disc, varies relatively slowly over time, and is likely accelerated by magnetic field lines anchored to the inner accretion disc that generate the magneto-rotational instability (MRI) and provide the viscosity though which angular momentum is transferred away from the accreting material. The rapid variability of the X-ray emission in this model, though, originates in a collimated core within the corona that is energised at its base. Fluctuations in luminosity propagate upwards through this core. Although I\,Zw\,1 is a radio-quiet AGN and does not exhibit a large-scale jet, this collimated core within the corona is reminiscent of a failed jet \citep{propagating_lag_paper}. In 2015, an X-ray flare was seen and was found to originate in the extended component of the corona, propagating inwards until it was finally observed in the emission from the collimated core \citep{1zw1_corona_paper}. The corona is accompanied by a multi-component, variable outflow, detected through absorption lines it imprints on the soft X-ray spectrum \citep{costantini+07}, and possibly found to vary over the course of the X-ray flare \citep{silva_1zw1}.

The first hard X-ray observations of I\,Zw\,1 made by \textit{NuSTAR} in 2020 show that the 3-50\keV\ X-ray spectrum is well described by the combination of the primary continuum emission from the corona and its reflection from the accretion disc around a spinning black hole ($a = J/M > 0.75\,GMc^{-2}$). The detection of reflected X-rays from the inner accretion disc is further corroborated by the measurement of reverberation time lags between the continuum and both the soft X-ray excess and the iron K line. The measured time lag corresponds to an X-ray emitting corona located a height of approximately 4\rg\ above the disc, where the gravitational radius $r_\mathrm{g} = GM/c^2$ is the radial coordinate of the event horizon in the equatorial plane about a maximally spinning black hole. During the 2020 observations, bright X-ray flares were observed, during which short flashes of redshifted iron K line emission were seen. These are consistent with the re-emergence of X-rays that were reflected from the back side of the disc, behind the black hole, and bent into our line of sight by the strong gravitational field, predicted when an X-ray flare is seen reverberating off the inner accretion disc \citep{1zw1_nature}.

We report on the 2020 observations of I\,Zw\,1 made with the \textit{NuSTAR} and \textit{XMM-Newton} X-ray observatories. We analyse the broadband X-ray spectrum and explore variability in the X-ray spectrum in the context of variations in the X-ray emitting corona over the course of the flare that shed light on the structure of the corona and the mechanisms by which it is powered and through which the flaring occurred.

\section{Observations and Data Reduction}
\begin{figure}
	\includegraphics[width=\columnwidth]{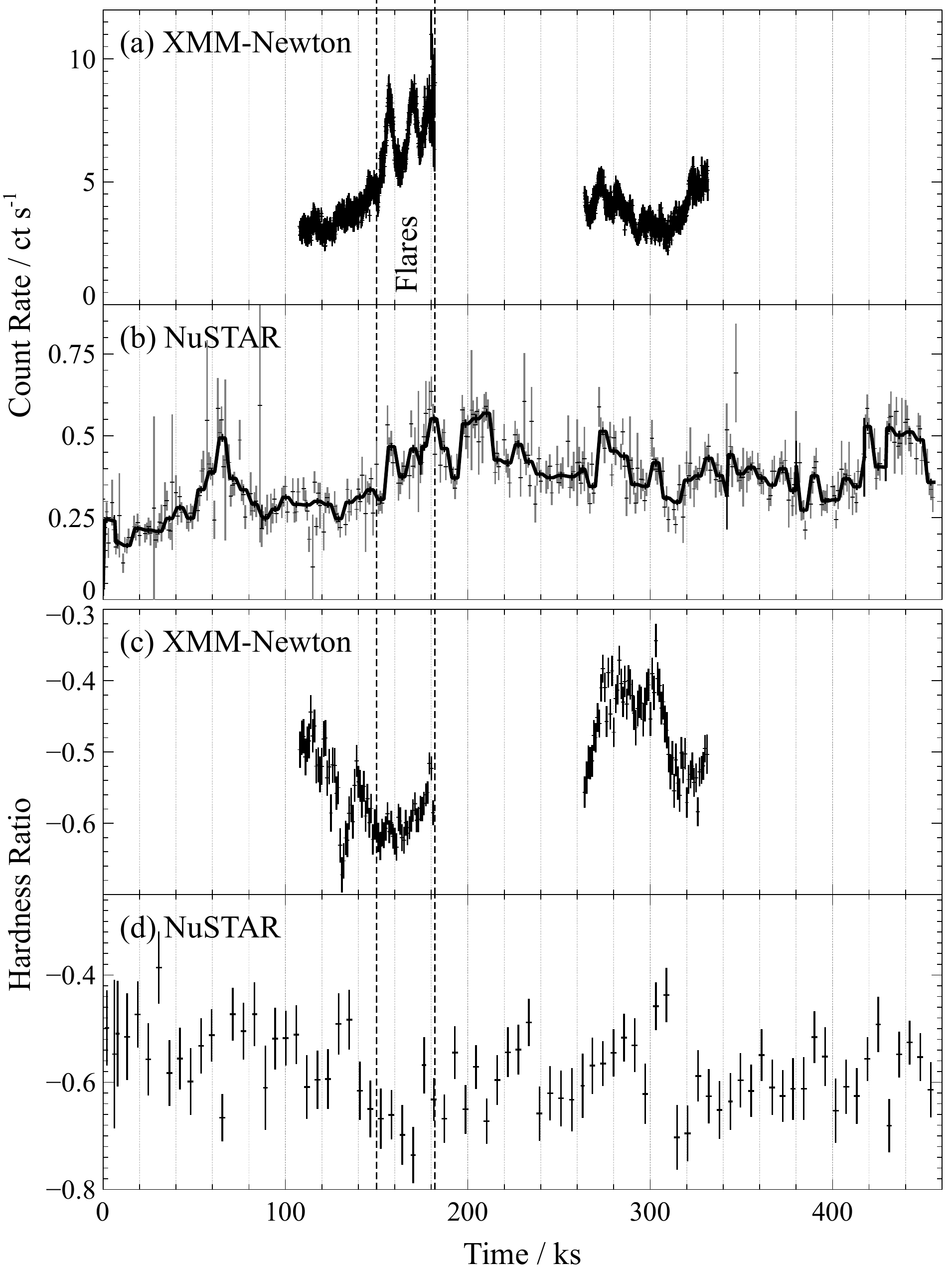}
	\caption{X-ray light curves of I\,Zw\,1 in 2020 January, (a) in the 0.3-10\keV\ band, measured with the \textit{XMM-Newton} EPIC pn camera, and (b) in the 3-50\keV\ band, measured by \textit{NuSTAR} and summed between the FPMA and FPMB detectors in 1000\s\ time bins. The overlaid line shows the light curve binned per spacecraft orbit. X-ray flaring was observed 150\ks\ from the beginning of the observations, toward the end of the first \textit{XMM-Newton} exposure, which had subsided by the time the second \textit{XMM-Newton} exposure had commenced. The lower panels show the hardness ratio (c) between the 0.3-1 and 2-10\keV\ bands, measured by \textit{XMM-Newton} in 1000\s\ time bins, and (d) between the 3-10 and 10-50\keV\ bands, measured by \textit{NuSTAR}, binned by spacecraft orbit. Vertical dashed lines indicate time time interval during which X-ray flaring was detected in the \textit{XMM-Newton} observations.}
	\label{fig:lc}
\end{figure}

I\,Zw\,1 was observed simultaneously by \textit{NuSTAR} \citep{nustar} and \textit{XMM-Newton} \citep{xmm} between 2020 January 11 and 2020 January 16. \textit{NuSTAR} obtained a total of 233\ks\ exposure over a continuous period of 5.3 days. \textit{XMM-Newton} observed I\,Zw\,1 twice during this period, obtaining continuous exposures of 76 and 69\,ks. The observations are outlined in Table~\ref{tab:data}, and the observed light curves are shown in Fig.~\ref{fig:lc}.

\begin{table}
		\caption{The \textit{NuSTAR} and \textit{XMM-Newton} observations of I\,Zw\,1 obtained in 2020 that are incorporated in this analysis.}
	\begin{center}		
		\begin{tabular}{lllc}
		\hline
		Observatory & OBSID & Start Date & Exposure \\
		\hline
		\textit{NuSTAR} & 60501030002 & 2020-01-11 & 233\ks \\
		\hline
		\textit{XMM-Newton} & 0851990101 & 2020-01-12 & 76\ks \\
		& 0851990201 & 2020-01-14 & 69\ks \\
		\hline
		\end{tabular}
	\end{center}
	\label{tab:data}
\end{table}

Toward the end of the first \textit{XMM-Newton} observation, 150\,ks from the start of the \textit{NuSTAR} observing period, flaring of the X-ray emission was observed. Two flares, most prominent in the soft X-ray band, were observed. Each lasted approximately 10\ks\ and the count rate peaked at 2.5 times the mean count rate that was seen before the flaring started. From the hard X-ray light curve obtained by \textit{NuSTAR}, the flaring can be seen to continue after the first \textit{XMM-Newton} observation ended, but had finished before the second \textit{XMM-Newton} observation began.

\subsection{\textit{NuSTAR} data reduction}
The \textit{NuSTAR} observations were reduced using the \textit{NuSTAR} data-analysis system, \textsc{nustardas}, v1.9.2. The event lists from each of the focal plane module (FPM) detectors were reprocessed and filtered using the \textsc{nupipeline} task, applying the most recent calibration available at the time of writing. We extracted the source photons from a circular region, 30\,arcsec in diameter, centered on the point source. We find that a smaller 30\,arcsec extraction region, suitable for X-ray sources that are fainter in the hard X-ray band, improves the signal to background ratio in the observations of I\,Zw\,1 compared to the larger 60\,arcsec region commonly employed for brighter targets. The background was extracted from a region, the same size, away from the point source, on each detector. Source and background spectra, along with their corresponding response matrices and ancillary responses (the effective area functions) were extracted using the \textsc{nuproducts} tasks. In addition, \textsc{nuproducts} was used to extract source and background light curves with all appropriate dead time and exposure corrections applied.

Initially, the separate spectra obtained from the FPMA and FPMB detectors were analysed separately. Fitting the spectra with a power law model, we find the best-fitting index to be consistent between the two detectors within statistical uncertainty. We therefore proceed to sum the spectra from FPMA and FPMB and analyse them using response matrices averaged between the two detectors, in order to maximise the signal-to-noise in the highest energy X-ray bands. Based upon the consistency between the FPMA and FPMB spectra, we expect that any uncertainty introduced by differences in the calibration and response between the two detectors and their respective telescopes to be dominated by statistical uncertainties arising from the limited photon count rate. I\,Zw\,1 was significantly detected by \textit{NuSTAR} above the background up to 50\,keV.

\subsection{\textit{XMM-Newton} data reduction}
We analyse primarily the data collected by the EPIC pn camera on board \textit{XMM-Newton}, due to the instrument's superior sensitivity, particularly when analysing the variability of the X-ray emission. During the observations, the pn camera was operated in small window mode so as not to be impacted by photon pile-up given the average count rate observed from I\,Zw\,1. The \textit{XMM-Newton} observations were reduced using the \textsc{xmm science analysis system (sas)} v18.0.0. The event lists were reprocessed using the \textsc{epproc} task, applying the latest available version of the calibration. Source photons were extracted from a circular region, 35\,arcsec in diameter, and the background was extracted from a circular region of the same size, located on the same detector chip. The source and background spectra were extracted using the \textsc{evselect} task and the corresponding response and ancillary response matrices were generated using \textsc{rmfgen} and \textsc{arfgen}. Light curves were extracted from the \textit{XMM-Newton} observations, also using \textsc{evselect}, and were corrected to account for dead time and exposure variations using the \textsc{epiclccorr} task.

\subsection{Assessing the X-ray variability}
To obtain an initial estimate of the spectral variability, we calculate the X-ray hardness ratio as a function of time. The hardness ratio, between the count rates in a hard X-ray band $H$ and soft X-ray band $S$, is defined as $(H-S)/(H+S)$. From the \textit{XMM-Newton} observations, we compute the hardness ratio between the 0.3-1 and 2-10\keV\ bands to assess variability in the shape of the soft X-ray spectrum. From the \textit{NuSTAR} observations, we compute the hardness ratio between the 3-10 and 10-50\keV\ bands, to assess variability in the shape of the hard X-ray spectrum.

The hardness ratios as a function of time are shown alongside the light curves in Fig.~\ref{fig:lc}. Substantial variability can be seen in the soft X-ray spectrum. Leading into the flares the X-ray spectrum can be seen to soften. At the beginning of the second \textit{XMM-Newton} observation, once the flares had subsided, the spectrum hardens significantly. On the other hand, variability in the shape of the hard X-ray spectrum, as seen by \textit{NuSTAR}, is much less. The hardness ratio in the \textit{NuSTAR} bands remains much more constant, though a slight softening of the spectrum can be seen during the flares.

In Section~\ref{sec:var}, we present a detailed analysis of the variability in the X-ray spectrum to determine its underlying causes, as well as the cause of the flares.

\section{Modelling the broadband X-ray Spectrum}
The 3-50\,keV X-ray spectrum of I\,Zw\,1 is well-described by continuum emission from the corona, possessing a power law spectrum, and the reflection of this continuum emission from the accretion disc \citep{1zw1_nature}. In addition, time lags between the continuum emission, and both the soft X-ray emission and iron K line, suggest the reverberation of continuum variations. We therefore begin with this model to describe the full 0.3-50\keV\ band covered by the combination of \textit{XMM-Newton} and \textit{NuSTAR}. The continuum emission and the reflection from the accretion disc are described by the \textsc{relxill} model \citep{dauser+15}, which includes reprocessing of the incident continuum emission by the plasma in the accretion disc, from the \textsc{xillver} model \citep{garcia+2010,garcia+2011,garcia+2013}, as well as the relativistic broadening of this reflection spectrum by Doppler shifts and gravitational redshifts from an accretion disc orbiting a spinning black hole. Reflection from the inner accretion disc contributes a relativistically broadened iron K line, a Compton hump, peaking between 20 and 30\keV, and a soft excess comprised of bremsstrahlung emission and soft X-ray emission lines that are blended together below around 1\keV. We here seek to extend the spectral model to the cover the full \textit{XMM-Newton} and \textit{NuSTAR} bandpass, from 0.3 to 50\keV.

The continuum model is described by the photon index (the slope of the power law) and the high energy cutoff, corresponding to the temperature of the corona, above which the power law spectrum transitions to an exponential cutoff. The model of the reflection from the accretion disc has parameters corresponding to:
\begin{itemize}
	\item The iron abundance in the disc, $A_\mathrm{Fe}$.
	\item The ionisation parameter, defined by the ratio of the ionising flux to the density, $\xi = 4\pi F/n$.
	\item The emissivity profile of the disc, defined as the reflected flux as a function of radius, measured in the rest frame of the reflecting material. The emissivity profile depends upon the location and geometry of the corona that illuminates the disc and is parametrised as a broken power law.
	\item The inclination, $i$, of the disc to the line of sight.
	\item The inner radius of the accretion disc, $r_\mathrm{in}$, from which the spin of the black hole can be inferred if it coincides with the innermost stable orbit, $r_\mathrm{ISCO}$.
	\item The reflection fraction, $R$, defining the observed ratio of the reflected to continuum flux (integrated over the 0.3-100\keV\ energy band).
\end{itemize}

The soft X-ray spectrum exhibits absorption, not just from the interstellar medium along the line of sight in our own Galaxy, but from a multi-component, variable outflow intrinsic to I\,Zw\,1 \citep{costantini+07}. We model the absorption using the \textsc{xstar} photoionisation code \citep{xstar}. Since the outflows in I\,Zw\,1 have been observed to vary in time, \textsc{xstar} was used to generate grids of absorption models that span the ranges of column densities and ionisation parameters measured in previous observations of I\,Zw\,1 \citep{silva_1zw1}. While the EPIC pn spectrum does not have the energy resolution required to resolve the absorption lines and accurately measure the column density, velocity and ionisation state of the outflows, these models account for the shape of the soft X-ray spectrum and ensure that uncertainty in the absorption is properly accounted for when estimating the parameters of the coronal X-ray emission and reflection from the accretion disc. We apply two multiplicative absorption models to the spectrum to describe the two outflow components that have been detected in I\,Zw\,1 \citep{silva_1zw1}, but allow the column density, ionisation parameters and redshifts (which account for the velocities of the components) to vary freely as required to describe the observed spectrum. We then marginalise over uncertainties in the absorption parameters when estimating the uncertainties associated with the continuum and reflection model parameters. Preliminary analysis of the high-resolution spectrum obtained by the \textit{XMM-Newton Reflection Grating Spectrometer (RGS)} in 2020 shows that the two warm absorber components are still present. The first component is measured with a column density of $1.3\times 10^{21}$\psqcm, an ionisation parameter of $\log (\xi / \mathrm{erg\,cm\,s}^{-1}) = -1.1$ and an outflow velocity of 1700\kmps, while the second component has a column density of $1.1\times 10^{21}$\psqcm, an ionisation parameter of $\log (\xi / \mathrm{erg\,cm\,s}^{-1}) = -2.4$ and an outflow velocity of 3200\kmps. For each component, the lines are broadened, corresponding to a turbulent velocity of 100\kmps. Although these measurements are made from the time-averaged spectrum across the entirety of the observations, the parameters of the warm absorbers measured by the RGS are roughly consistent with the constraints on the warm absorbers obtained from the lower resolution EPIC pn spectra. It should be noted, however, that the EPIC pn spectra resolve only the shape of the absorbed continuum and the most significant edges and cannot resolve the narrow absorption lines that enable detailed measurement of the absorbers with the RGS.

Due to the high degree of spectral variability we observe via the hardness ratio, we divide the observations into three time periods: before the flaring begins, the flares, and the period after the flares. The model was fit simultaneously to the spectra from the three time intervals. The spin of the black hole (and inner radius of the accretion disc), the inclination of the accretion disc to the line of sight and the iron abundance in the disc were tied between the three intervals, since these parameters should not have changed over the course of the observations. We allow all other parameters including the ionisation of the disc, the location and geometry of the corona (encoded in the emissivity profile of the reflection component observed from the disc), the reflection fraction, and the column densities, ionisation states and velocities of the warm absorbers to vary between the three time periods, should the data require these parameters to vary. Fitting the spectra from the three time periods simultaneously in this way determines not only the best-fitting model, but requires that the model reproduces the observed spectral variability in a physically meaningful manner.

We fit the model simultaneously to the \textit{XMM-Newton} spectra over the 0.3-10\keV\ energy band and the \textit{NuSTAR} spectra over the 3-50\keV\ band during the three time periods using \textsc{xspec} \citep{xspec}. The model applied to the \textit{NuSTAR} spectrum is multiplied by a constant to account for calibration uncertainties between the instruments. We find the best-fitting value of the constant, representing the normalisation of the \textit{NuSTAR} spectrum with respect to the \textit{XMM-Newton} spectrum, to be $1.06_{-0.02}^{+0.03}$. The value of this cross-calibration constant was tied between the three time intervals.

The model parameters were optimised to minimise the modified $C$-statistic, which is based upon the Cash statistic \citep{cash} but modified such that when the number of counts is high, $C$ tends toward the $\chi^2$ statistic.\footnote{\url{https://heasarc.gsfc.nasa.gov/xanadu/xspec/manual/XSappendixStatistics.html}} Once the best-fitting parameters have been found, we estimate the parameter uncertainties using a Markov chain Monte Carlo (MCMC) calculation to explore the parameter space. MCMC chains were run using the algorithm of \citet{goodman_weare}, implemented in \textsc{xspec}. Chains were run with 80 walkers for $10^7$ iterations, discarding (or `burning') the first 5000 iterations so as to remove any `memory' of their starting positions.

\begin{figure}
	\includegraphics[width=\columnwidth]{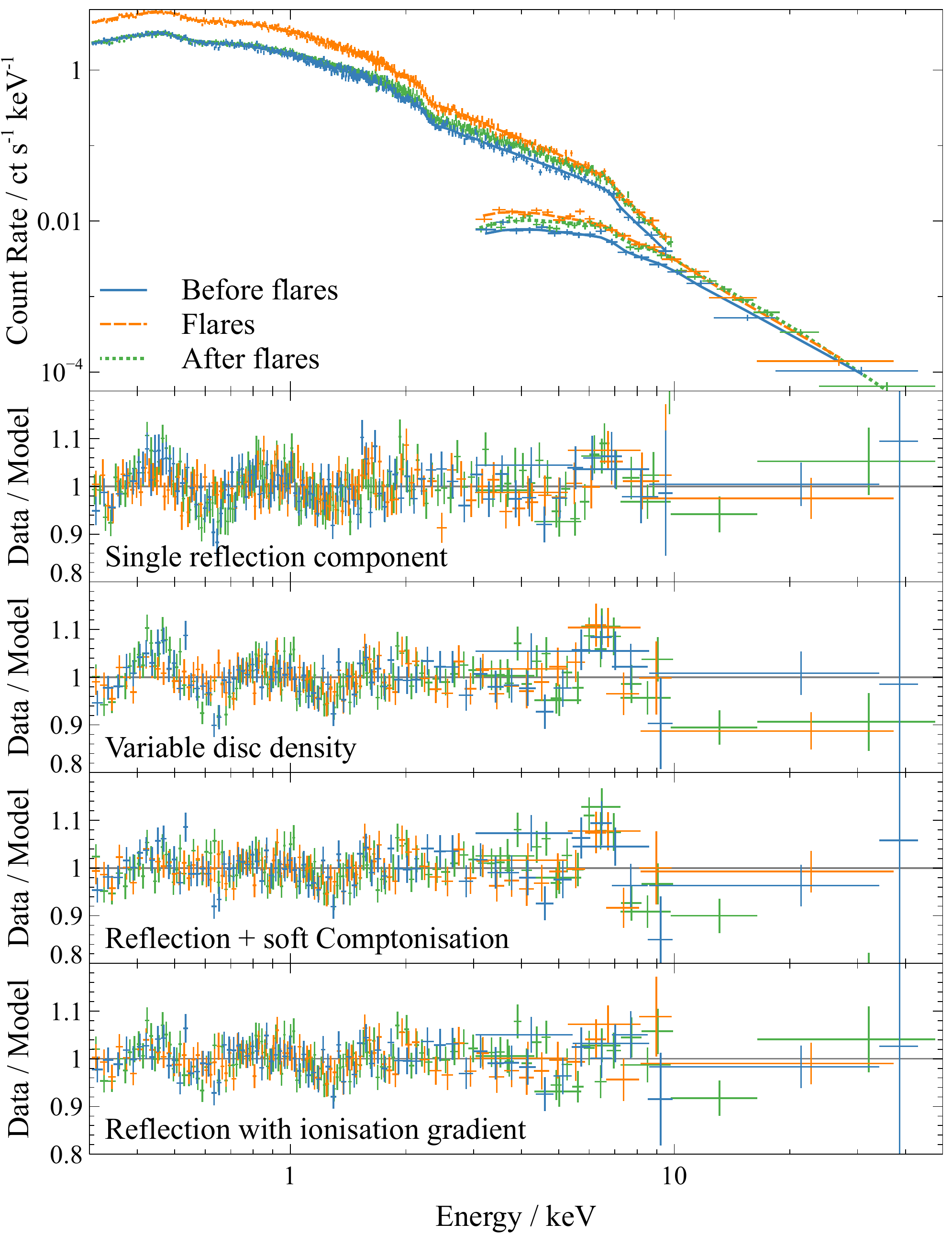}
	\caption{The 0.3-50\keV\ X-ray spectrum of I\,Zw\,1, measured by \textit{XMM-Newton} and \textit{NuSTAR}. The time-averaged spectrum over the whole of the observations is shown. The top panel shows the observed spectrum with the best-fitting model, consisting of the continuum, reflection from the accretion disc with a radial gradient in the ionisation parameter, and two warm absorption components. The lower panels show the ratio between the observed spectrum and four candidate models: a single, fixed-density reflection component, a single reflection component with variable density, a reflection component with additional soft excess component arising from warm Comptonization, and a  reflection component with a radial ionization profile across the disc. Each of these models includes the continuum emission from the corona and the warm absorbers.}
	\label{fig:spectrum}
\end{figure}

We find that although the initial model of the continuum, reflection from the accretion disc, and absorption by two outflow components, provides a good description of the 3-50\keV\ spectrum, this model provides only a reasonable description of the spectrum over the full 0.3-50\keV\ band (Fig.~\ref{fig:spectrum}), yielding a $\chi^2$ statistic of 3380 for 3161 degrees of freedom ($\chi^2 / \nu = 1.07$). The residuals between the model and the observed spectrum show that this simple model cannot simultaneously account for the shape of the soft excess and the amplitude of the iron K line. We therefore assess modifications to the model of reflection from the accretion disc.

In order to compare the descriptions of the observed spectrum provided by different models, we employ the deviance information criterion, DIC \citep{dic}. DIC is a Bayesian model selection metric that factors in both the goodness of fit (from the likelihood function) and the number of free parameters. The DIC is a hierarchical generalization of the Akaike information criterion (AIC), which estimates the relative quality of a set of models in representing a specific set of data (in this case the observed spectrum), in terms of the information lost by each model. The DIC statistic is an asymptotic approximation to the model evidence, \textit{i.e.} the probability that a given model will produce the observed data, integrated over the entire parameter space. When simply comparing fit statistics, such as the minimum $\chi^2$ obtained between models, it is possible for a model to not only under-fit the data (thereby providing a poor fit), but also to over-fit the data by having too many free parameters. AIC and DIC statistics address this problem by penalising models with more free parameters, and represent a trade-off between the goodness of fit and the simplicity of each model. DIC can be computed directly from the results of MCMC calculations and has the advantage over other Bayesian model selection techniques and information criteria (\textit{e.g.} the BIC or AIC) of defining an effective number of parameters. The DIC thereby does not penalise a model for `nuisance' parameters that are not constrained by the data. In general, a model yielding a lower DIC is preferred by the data, and the $\Delta$DIC between models quantifies the strength of the evidence in favour of one model over another. $\Delta$DIC values between zero and two suggest only marginal evidence, while $\Delta$DIC greater than 6 shows strong evidence for one model in favour of another \citep{kass+1995}.

The mismatch between the shape of the soft excess and the iron K line could arise due to the accretion disc density being greater than the canonically assumed electron density of $n_\mathrm{e} = 10^{15}$\pcmcu. Increasing the density of the accretion disc leads to the trapping of radiation in the upper layers, increasing the temperature and increasing the soft X-ray emission through bremsstrahlung \citep{xillver_density}. We therefore use the \textsc{relxilld} model to describe the continuum and reflection in place of \textsc{relxill}. \textsc{relxilld} models the reflection from an accretion disc with variable electron density. We find that this model improves the description of the soft X-ray spectrum, with $\Delta\chi^2 = -10$ compared to the fixed-density reflection model, with one fewer degree of freedom. The improvement of the fit is not enough, however, to offset the additional free parameter in this model, and is slightly disfavoured by the DIC statistic with $\Delta\mathrm{DIC} = +6$. The accretion disc in I\,Zw\,1 is best-described with a slightly-enhanced electron density of $\log(n_\mathrm{e}\,/\,\mathrm{cm}^3) = 16.7_{-1.1}^{+0.3}$.

It possible that the soft X-ray excess arises from a separate emission component from a `warm corona' that extends over the inner regions of the accretion disc \citep[\textit{e.g.}][]{done_jin,petrucci+2018}. In this model, the energy liberated at small radii in the disc is not fully thermalised, but is split between both an optically thin corona (which produces the high energy X-ray continuum with a power-law spectrum), and a lower temperature ($\sim0.2$\keV) optically-thick corona. Like the optically thin corona, this optically thick corona Comptonises the thermal photons emitted from the accretion disc, but instead of producing a power law spectrum extending to high energies, the warm corona produces an excess of soft X-ray emission, peaking below 1\keV. To test this model against the observed spectra of I\,Zw\,1, we replace the power law continuum component of the spectral model with \textsc{optxagnf}, an energetically self-consistent model which includes the power law continuum from the optically thin corona and the soft excess from the warm corona. We also include the reflection from the accretion disc, modelled by \textsc{relxill}, to account for the broad iron K line and the Compton hump, and the warm absorbers. We fit the radius of the warm corona, its temperature and optical depth to the observed spectrum, and allow these parameters to vary between the three time intervals. It is necessary to apply a high-energy cutoff to the power law continuum spectrum in order to match the observed \textit{NuSTAR} spectra. We find that while adding the additional soft excess component provides a much better description of the soft X-ray spectrum below 1\,keV, providing $\Delta$DIC$=-50$ compared with the original model, this model is still unable to simultaneously account for the shape of the soft excess and the strength of the observed iron K line. Broad residuals are still seen around 6\keV\ in the spectrum. Indeed, since we found that the simplest reflection model alone \textit{underestimates} the iron K line with respect to the soft excess, adding further emission components to the soft X-ray band does not remedy this problem.

The shape of the soft excess can also be altered by the ionisation structure of the accretion disc. If the accretion disc is photoionised by the irradiating flux from the corona, we expect that the ionisation state of the disc varies as a function of radius. The ionisation parameter is defined $\xi = 4\pi F / n$ for a plasma with number density $n$, receiving ionising flux $F$. When the disc is illuminated by a compact corona close to the black hole, light bending toward the black hole focuses the irradiating flux onto the inner parts of the disc, causing it to fall off as steeply as $r^{-7}$ over the inner few gravitational radii, then following approximately $r^{-3}$ over the outer disc. On the other hand, the density of a thin accretion disc is expected to decrease as $n \propto r^{-\frac{3}{2}}$ in regions where the gas pressure dominates, though increasing with radius as $n \propto r^\frac{3}{2}$ in the innermost regions where radiation pressure becomes dominant \citep{novthorne}. Depending upon the precise density profile of the disc, the ionisation parameter can fall off as steeply as $r^{-\frac{17}{2}}$ in the innermost regions, tending to $r^{-\frac{3}{2}}$ over the outer disc \citep[see \textit{e.g.}][]{plunging_region_paper}.

The \textsc{relxilllpion} variant of the \textsc{relxill} model approximates the variation in ionisation parameter across the accretion disc as a single power law, the index of which is fit to the data. We find that allowing the ionisation of the accretion disc to vary as a function of radius in this way, significantly improves the description of the observed spectrum, and simultaneously fits the soft X-ray spectrum, the iron K line and the Compton hump, providing $\Delta\mathrm{DIC} = -110$ compared to the original, fixed-density, constant-ionisation reflection model, and $\Delta\mathrm{DIC} = -60$ with respect to the model with an additional soft X-ray emission component arising from warm Comptonisation. Before and after the flares, the ionsiation of the disc is relatively low, with $\log(\xi_\mathrm{in} / \mathrm{erg\,cm\,s}^{-1}) \sim 1$. The ionisation parameter during this time periods was found to fall off as $\sim r^{-\frac{1}{2}}$. During the flares, the ionisation parameter is not well constrained, but a relatively steeply falling profile with power law index greater than three is required to fit the spectrum. It should be noted that the power law index is expected to vary as a function of radius and the best-fitting index represents a flux-weighted average over the entire disc. In \textsc{relxilllpion}, the density of the accretion disc is frozen at the canonical value of $n_\mathrm{e} = 10^{15}$\pcmcu.

\begin{figure}
	\includegraphics[width=\columnwidth]{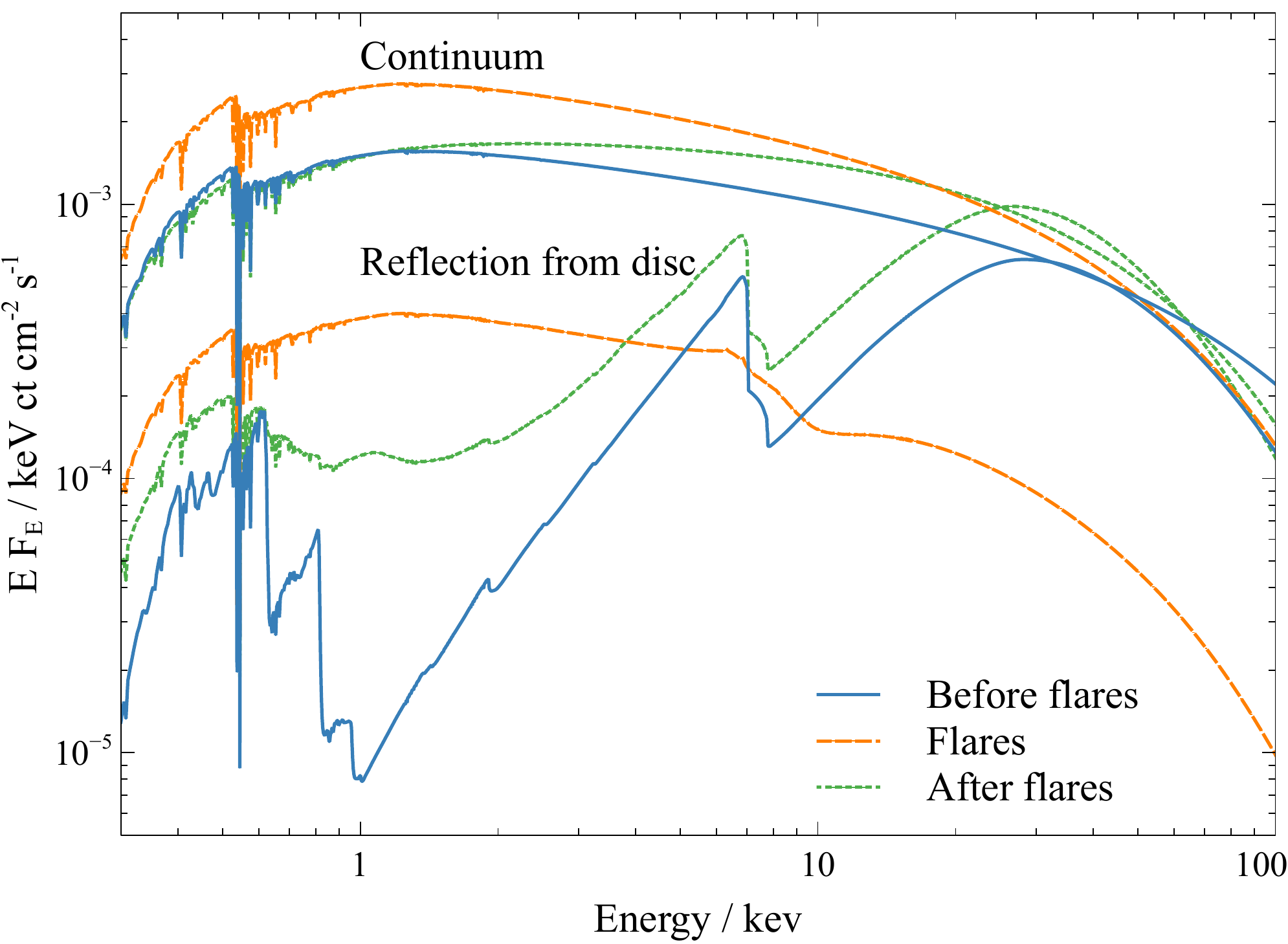}
	\caption{The best-fitting model to the 0.3-50\keV\ X-ray spectrum of I\,Zw\,1, fit simultaneously to the observations in three time intervals: before, during and after the flares. The model consists of the continuum emission from the corona (upper lines), relativistically-broadened reflection component from the accretion disc with radial ionisation gradient approximated by a power law (lower lines), and warm absorption from two outflow components (applied to each of the continuum and reflection components).}
	\label{fig:model}
\end{figure}

We conclude that the model best-describing the observed spectra consists of the directly-observed continuum emission from the corona, as well as reflection from an accretion disc with a radial ionisation gradient, in addition to the two warm absorption components that were previously discovered in I\,Zw\,1. This model provides an adequate description of the data, yielding a $\chi^2$ statistic of 3253 for 3161 degrees of freedom ($\chi^2 / \nu = 1.03$). The best-fitting parameters to the time-averaged spectrum obtained from the full observations are shown in Table~\ref{tab:param} and the model components are shown in Fig~\ref{fig:model}.

\subsection{Ultra-fast outflows}
\citet{reeves_braito_2019} report the detection of an ultra-fast outflow (UFO) in previous observations of I\,Zw\,1 via the presence of P Cygni-like emission and absorption features in the iron K band of the X-ray spectrum. The presence of blueshifted absorption at 9\keV\ and excess emission around 7\keV\ suggests a highly ionised wind, launched from the accretion disc at a velocity of at least $0.25c$. To determine if this wind was still present in the 2020 observations, we search for similar features in the X-ray spectrum by adding a narrow Gaussian emission or absorption line to the best-fitting model spectrum. The width of the line was frozen at 0.01\keV\ (\textit{i.e.} below the energy resolution of the detectors so as to search for unresolved emission and absorption lines) and the line was scanned through the spectrum to determine the improvement in the $\chi^2$ fit statistic as a function of line energy and normalisation (see \citealt{parker_iras_nature} for a discussion of the method).

\begin{figure}
	\includegraphics[width=\columnwidth]{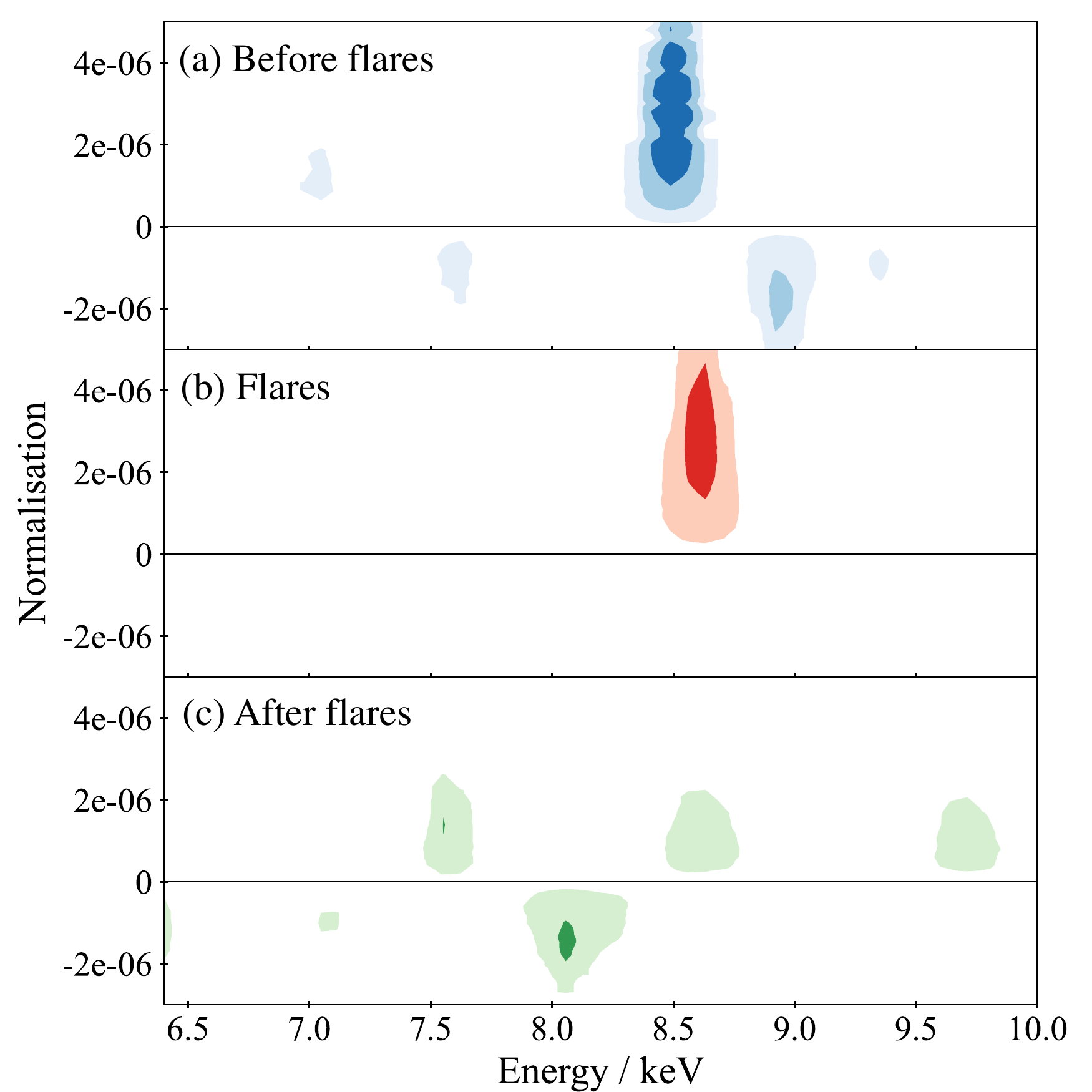}
	\caption{The improvement in the $\chi^2$ fit statistic when narrow Gaussian absorption and emission features are added at different energies in the iron K band, to search for signatures of ultra-fast outflows (UFOs). The model is applied simultaneously to the \textit{XMM-Newton} and \textit{NuSTAR} spectrum. Positive normalisations correspond to emission features, negative normalisation to absorption features. The detected features are shown from time intervals (a) before, (b) during, and (c) after the flares. Contours correspond to 1, 2 and $3\sigma$ detection significance.}
	\label{fig:linescan}
\end{figure}

The improvement in fit statistic with the addition of narrow emission and absorption features in time periods before, during and after the flares, is shown in Fig~\ref{fig:linescan}. We find similar absorption and emission features that could be attributed to an ultra-fast outflow, though in the 2020 observations we find that the absorption features are detected at only $2.7\sigma$ significance. The emission feature is most strongly detected before the flares, at $3.7\sigma$ significance. Interestingly, however, we note that the features in the spectrum attributable to the UFO vary. Before the flares, blueshifted absorption is detected at 9\keV, with an associated emission feature at 8.5\keV. During the flares, only the emission component is detected. After the flares, the absorption feature is detected, but at a lower energy of 8\keV, along with an associated emission feature at 7.5\keV. These results suggest that the flaring in the X-ray emission is accompanied by variations in the velocity and ionisation state of the UFOs launched from the inner disc, however there is insufficient signal to analyse this variability further with the present observations.

We find that in addition to only being marginally detected, including these Gaussian emission and absorption features in the model does not alter the best-fitting parameters of the coronal X-ray emission and the reflection from the accretion disc. We therefore do not consider these features further in this work.

\subsection{Constraining the geometry of the corona}
\label{sec:emissivity}
The emissivity profile of the disc, that is the variation in reflected flux as a function of radius, corresponds to the flux received from the corona at each radius on the disc. The \textsc{relxilllpion} approximates the corona as a point source located on the rotation axis above the black hole. Under this approximation, the height of the corona above the disc plane can be estimated to be $h=2.3_{-0.2}^{+1.2}$\rg, so as to provide the emissivity profile of the disc that best matches the observed reflection spectrum. Measuring the emissivity profile of the accretion disc directly, however, allows for the measurement of the location, geometry and spatial extent of the corona over the disc \citep{understanding_emis_paper}.

\begin{figure}
	\includegraphics[width=\columnwidth]{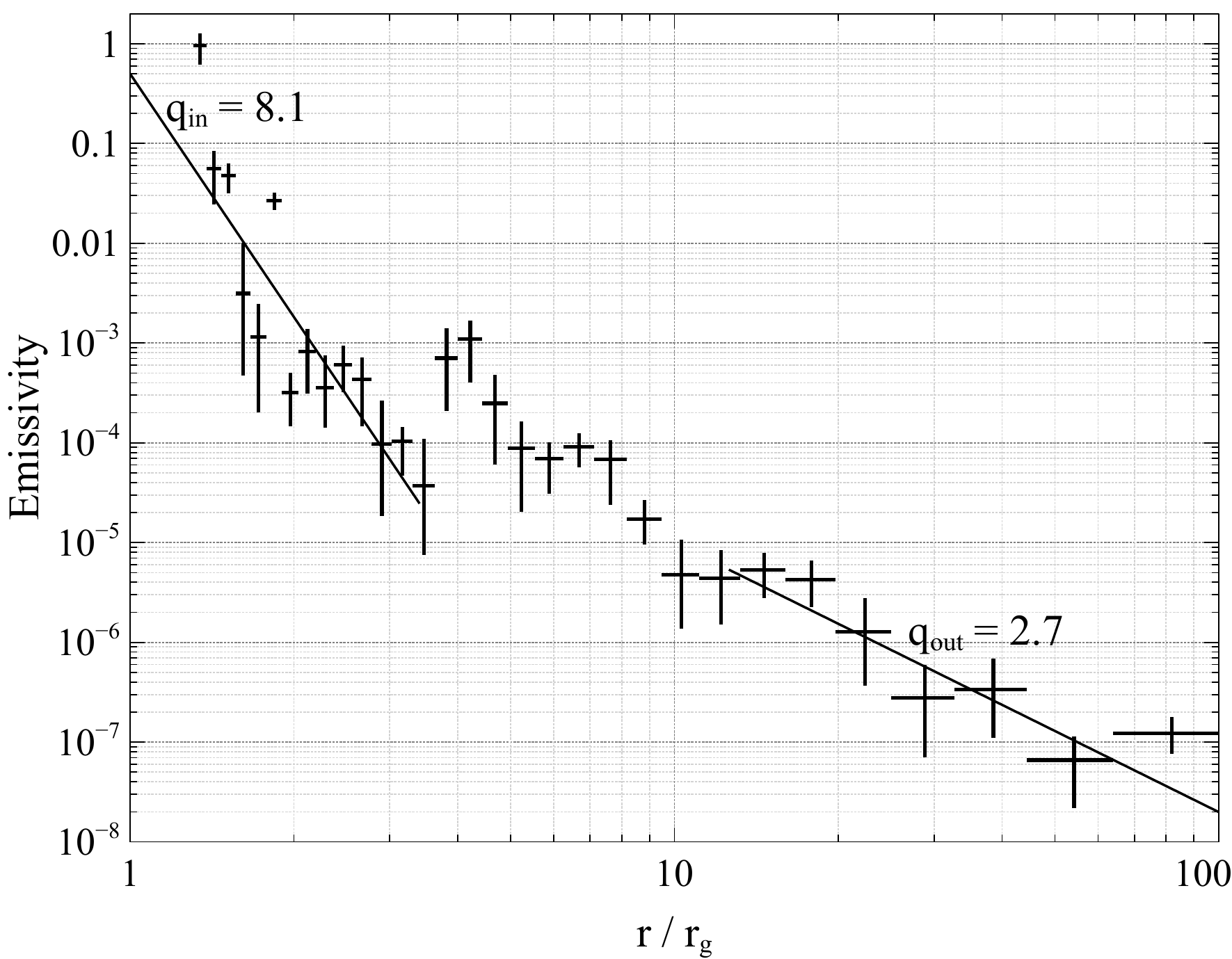}
	\caption{The emissivity profile of the accretion disc, defined as the reflected flux from the accretion disc as a function of radius (defined in the rest frame of the emitting material). The emissivity profile was measured by fitting the time-averaged reflection spectrum in the 3-7\keV\ range (\textit{i.e.} the iron K line), as the sum of the contributions from different radii. The emissivity profile shows the typical twice-broken power law form expected when the disc is illuminated by a compact corona. Solid lines show the best fitting power law indices to the inner and outer portions of the profile. The outer break radius corresponds to the extent of the corona over the accretion disc.}
	\label{fig:emis}
\end{figure}

We therefore measured the emissivity profile of the accretion disc from the observed shape of the relativistically broadened iron K line in the spectrum. Varying the relative illumination of the inner and outer disc varies the relative contribution of redshifted line photons to the observed spectrum and hence alters the shape of the redshifted wing of the line. We measured the emissivity profile fitting the reflection spectrum in the 3-7\keV\ range (\textit{i.e.} the iron K line), as the sum of the contributions from different radii \citep{1h0707_emis_paper}. The flux received from each radius was fit to the spectrum as a free parameter in an MCMC calculation. In order to measure the emissivity profile in this way, it is necessary to have detected sufficient photon counts from the broad iron K line so as to accurately constrain the contribution from each radius on the disc. We therefore obtain a first estimate of the emissivity profile by fitting the shape of the line in the time-averaged spectra across the whole period of the observations, though noting that variability in both the broad line and underlying continuum during this period could skew the measured profile. The measured emissivity profile is shown in Fig.~\ref{fig:emis}.

The emissivity is defined as the flux measured in the rest frame of the orbiting material in the disc and can be described by a twice-broken power law. Over the outer disc, the emissivity falls as $r^{-3}$, as expected for the illumination of a disc by a central source in flat spacetime, with no gravity. The emissivity of the inner disc falls steeply (following $r^{-8}$). The combination of light bending, focusing X-rays towards the black hole and hence onto the inner disc, with the blueshifting of rays as they travel closer to the black hole, strongly enhances the flux received by observers on the inner disc \citep{understanding_emis_paper}.

The emissivity profile shows evidence of flattening over the middle region. Beyond 3\rg\, the measured emissivity points lie above the expectation of the $r^{-8}$ power law observed over the innermost radii of the disc, and at around 8\rg\ and 20\rg, there is evidence of deviation from the $\sim r^{-3}$ decrease that describes the emissivity of the outer disc. Flattening of the emissivity profile to a power law index less than 3 can be understood in the context of a point source, where the flux received is approximately constant for $r \ll h$. For compact, point-like coron\ae\ higher above the black hole, the break radius from the flat middle portion of the profile to the $r^{-3}$ power law over the outer disc corresponds to the height of the corona. As the height increases, however, less illumination is received by the inner disc, so as the break radius moves out to larger radius, the steepness of the inner emissivity profile is expected to decrease. The simultaneous observation of an extended flat portion of the emissivity profile with a steep inner emissivity profile cannot be explained by a compact, point-like corona. The profile we observe can be explained by a corona at low height, but with a portion that is spatially extended over the surface of the disc. In this instance, the break radius corresponds to radial extent of the corona over the disc.

\subsection{Variability of the emissivity profile and coronal geometry}
If the geometry of the corona is variable over the time period of the observations, the emissivity profile we measure will be a superposition of the emissivity profiles at different times during the observation, where the outer break radius can move if the spatial extent of the corona over the accretion disc is changing. It is also possible that variation in the slope of the underlying X-ray continuum could be manifested as inaccuracies in the measured emissivity profile. We therefore measure the emissivity profile of the accretion disc during the three time periods, before, during and after the flares. These shorter time intervals do not have the required signal-to-noise to measure the emissivity profile as a free function of radius, however we are motivated by the approximate form of the emissivity profile measured from the time-averaged spectrum to fit the emissivity profile during each time period as a twice-broken power law, applying the \textsc{kdblur3}\footnote{\url{https://github.com/wilkinsdr/kdblur3}} relativistic blurring kernel to the \textsc{xillver} reflection model.

We find that the twice-broken power law model of the accretion disc emissivity profile is preferred over a model in which the emissivity profile is modelled by a single power law. The twice-broken power law model yields $\Delta$DIC$=-12$ over the single power law model. During each of the time intervals, the index over the innermost part of the disc is constrained to be greater than 7 at the 90 per cent confidence limit, showing that even accounting for spectral variability between the time periods, the observed iron K line supports the focusing of the X-ray continuum emission towards the black hole and onto the innermost parts of the disc, as predicted when reflection is observed from an accretion disc is illuminated by a compact corona.

The best-fitting value of the outer break radius of the emissivity profile was found to be $7_{-2}^{+4}$\rg\ before the flares and $6_{-2}^{+2}$\rg\ after the flares subsided. During the flares, there is insufficient exposure to tightly constrain the outer break radius, but the measured value of $18_{-7}^{+6}$ is suggestive that the corona expanded during the flares. Variation of the outer break radius can be explained either by a compact corona that varies in height above the accretion disc, or by the variation of the radial extent of the corona over the inner part of the disc \citep{understanding_emis_paper}. If the height of the corona above the disc increases, the increasing outer break radius of the emissivity profile is expected to be accompanied by a decrease in the slope of the inner part of the emissivity profile. The fact that the power law index over the innermost part of the disc is constrained to be $>7$ at all times suggests that it is the radial extent of the corona over the accretion disc that varies. The corona expands from a radius of $7_{-2}^{+4}$\rg\ before the flares, to $18_{-7}^{+6}$ during the flares, collapsing again to a radius of $6_{-2}^{+2}$\rg\ once the flares subside. The increase in the outer break radius between the before flare and flaring periods is detected only tentatively at 86.3 per cent ($1.5\sigma$) significance. The contraction after the flares is detected at 92.5 per cent ($1.8\sigma$) significance. Neither change in the outer break radius is detected to sufficient significance to be confident of variations in the coronal geometry, however hints of variation to the outer break radius might well be useful interpreting other aspects of the observed variability in the context of the corona.

\begin{figure}
	\includegraphics[width=\columnwidth]{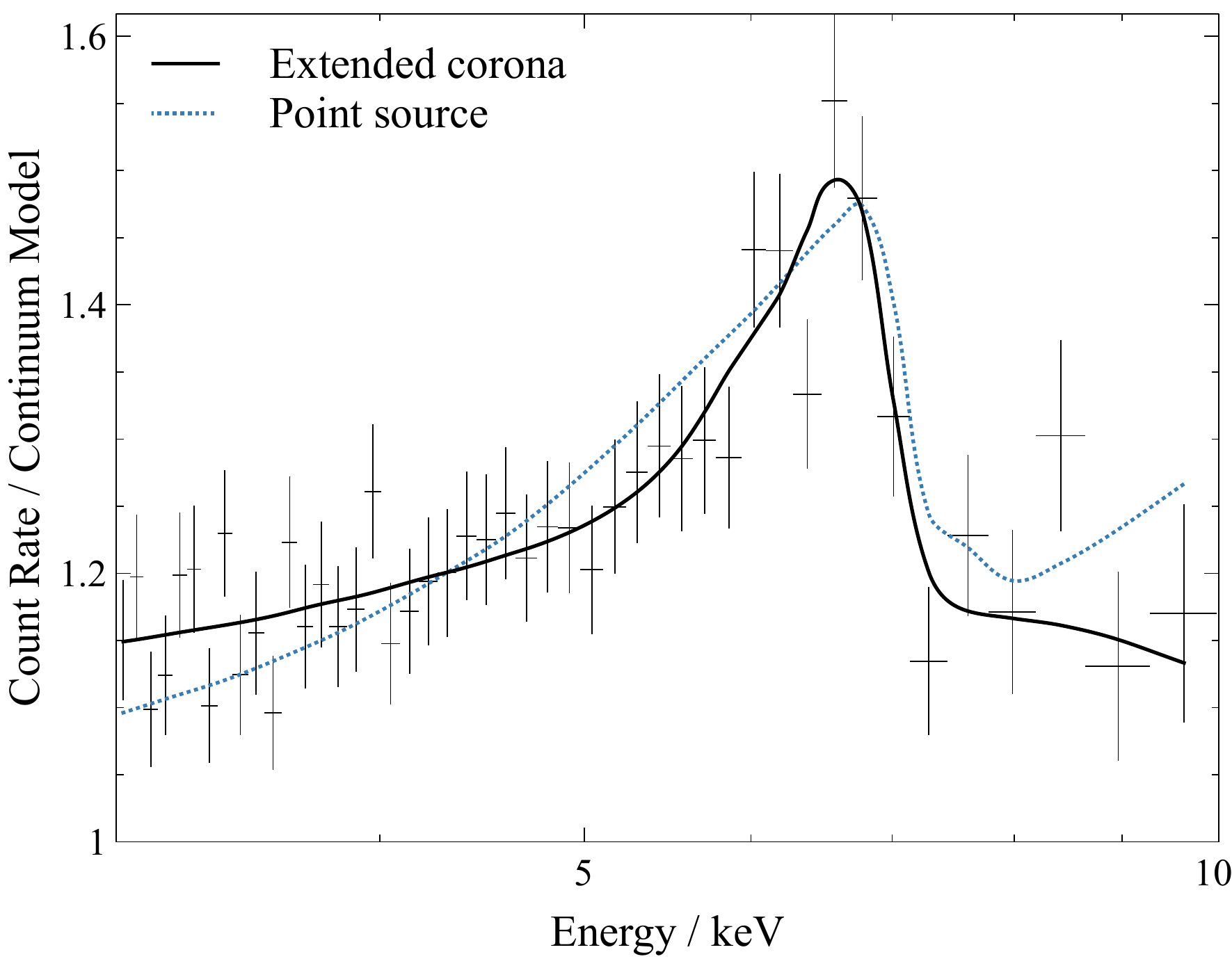}
	\caption{The ratio of the 3-10\keV\ band of the time-averaged X-ray spectrum, measured by \textit{XMM-Newton}, to the best-fitting power-law continuum component of the model. The ratio shows the profile of the relativistically broadened iron K line. The measured line profile is compared to models in which the emissivity profile of the accretion disc corresponds to illumination by a point source (\textsc{relxillionlp}), shown by the dashed line, and in which the emissivity profile is modelled by a twice-broken power law to account for the spatial extent of the corona (\textsc{kdblur3}), shown by the solid line. A model in which the corona is spatially extended is able to better-reproduce the shape of the redshifted wing of the line.}
	\label{fig:fek_model}
\end{figure}

Fig.~\ref{fig:fek_model} illustrates how the profile of the relativistically broadened iron K line from the inner accretion disc constrains the geometry of the corona. The observed line profile is compared to models in which the emissivity profile of the accretion disc corresponds to illumination by a point source on the rotation axis of the black hole, and in which the emissivity profile is given by the best-fitting twice-broken power law, accounting for the spatial extent of the corona over the disc. The twice-broken power law is able to better explain the detailed shape of the redshifted wing of the line.

\begin{table*}
	\caption{The best-fitting model parameters. The model was applied simultaneously to the spectra obtained in time intervals before, during and after the flares in the 0.3-10\keV\ 3-50\keV\ bands measured by \textit{XMM-Newton} and \textit{NuSTAR}, respectively. The continuum emission from the corona is described by a power law with an exponential cutoff at high energy. The reflection from the accretion disc accounts for a radial varation of the ionization parameter, described by a power law. The reflection spectrum was initially modelled assuming a point-like corona on the spin axis of the black hole at variable height, defining the initial estimate of the emissivity profile. The emissivity profile was then measured in detail and fit using a twice-broken power law. The soft X-ray spectrum is modified by two warm absorption components from outflows. Parameter values represent the maximum of the likelihood function and uncertainties and upper/lower limits correspond to the 90 per cent confidence interval derived from MCMC calculations. Parameter values in the left-most column were tied between the three time periods. Due to the limited exposure available during the flares, the outer break radius of the emissivity profile was fit to the combined spectrum before and during the flares to better-constrain the variation of this parameter.}
	\begin{tabular}{ll|c|ccc}
	\hline
	& &  & \textbf{Before flares} & \textbf{Flares} & \textbf{After flares} \\
	\hline
	Continuum & Photon index, $\Gamma$ &  & $2.22_{-0.03}^{+0.04}$ & $2.29_{-0.02}^{+0.06}$ & $2.06_{-0.04}^{+0.03}$ \\
	&	 cutoff energy, $E_\mathrm{cut}$ / keV &  & $140_{-20}^{+100}$ & $45_{-9}^{+40}$ & $70_{-30}^{+40}$ \\
	\hline
	Reflection	&	Black hole spin, $a / GM\,c^{-2}$ & $>0.94$ \\
	&	Inclination, $i$ $/$ deg & $52_{-4}^{+2}$ \\
	&	Iron abundance, $A_\mathrm{Fe}$ $/$ Solar & $4.9_{-0.9}^{+2.1}$ \\
	&	Ionisation at inner edge, $\log(\xi_\mathrm{in} / \mathrm{erg\,cm\,s}^{-1})$ & & $1.0_{-0.4}^{+0.1}$ & $<4$ & $1.1_{-3}^{+0.5}$ \\
	&	Ionisation power law index &  & $0.6_{-0.3}^{+0.2}$ & $>3$ & $0.5_{-0.3}^{+1.0}$ \\
	& Reflection fraction &  & $0.23_{-0.11}^{+0.04}$ & $0.07_{-0.04}^{+0.02}$ & $0.45_{-0.15}^{+0.03}$ \\
	\hline
	Point source approx. & Corona height, $h / r_\mathrm{g}$ &  & $3.7_{-0.4}^{+1.0}$ & $>14$ & $3.8_{-0.4}^{+0.6}$ \\
	\hline
	Emissivity profile & Inner index, $q_\mathrm{in}$ & & $>7$ & $>7$ & $>7$ \\
	& 	Inner break radius, $r_\mathrm{break,in}$ $/$ \rg & & $3.6_{-1.1}^{+0.4}$ & $3.5_{-1.3}^{+0.4}$ & $3.5_{-0.6}^{+0.5}$ \\
	&	Middle index, $q_\mathrm{mid}$ & & $<0.1$ & $<0.1$ & $<0.1$ \\
	&	Outer break radius,  $r_\mathrm{break,out}$ $/$ \rg &  & $7_{-2}^{+4}$ & $18_{-7}^{+6}$ &  $6_{-2}^{+2}$ \\
	&	Outer index, $q_\mathrm{out}$ & & $2.3_{-0.2}^{+1.6}$ & $3.3_{-1.1}^{+0.7}$ & $3.0_{-0.8}^{+0.9}$\\
	\hline
	Warm absorbers & Column density, $n_{\mathrm{H},1} / 10^{20}\,\mathrm{cm}^{-2}$ & & $5_{-1}^{+2}$ & $6.9_{-0.6}^{+1.7}$ & $4.6_{-2.0}^{+0.7}$ \\
	&	Ionisation, $\log(\xi_1 / \mathrm{erg\,cm\,s}^{-1})$ & & $-0.8_{-0.1}^{+0.3}$ & $-0.8_{-0.1}^{+0.2}$ & $-0.7_{-0.3}^{+0.2}$ \\
	&	Column density, $n_{\mathrm{H},2} / 10^{20}\,\mathrm{cm}^{-2}$ & & $<1.7$ & $<1.7$ & $1.7_{-0.5}^{+1.2}$ \\
	&	Ionisation, $\log(\xi_2 / \mathrm{erg\,cm\,s}^{-1})$ & & $<1$ & $<1.4$ & $<1$ \\
	\hline
	\end{tabular}
	\label{tab:param}
\end{table*}

\section{Variability during the X-ray flares}
\label{sec:var}
\begin{figure}
	\includegraphics[width=\columnwidth]{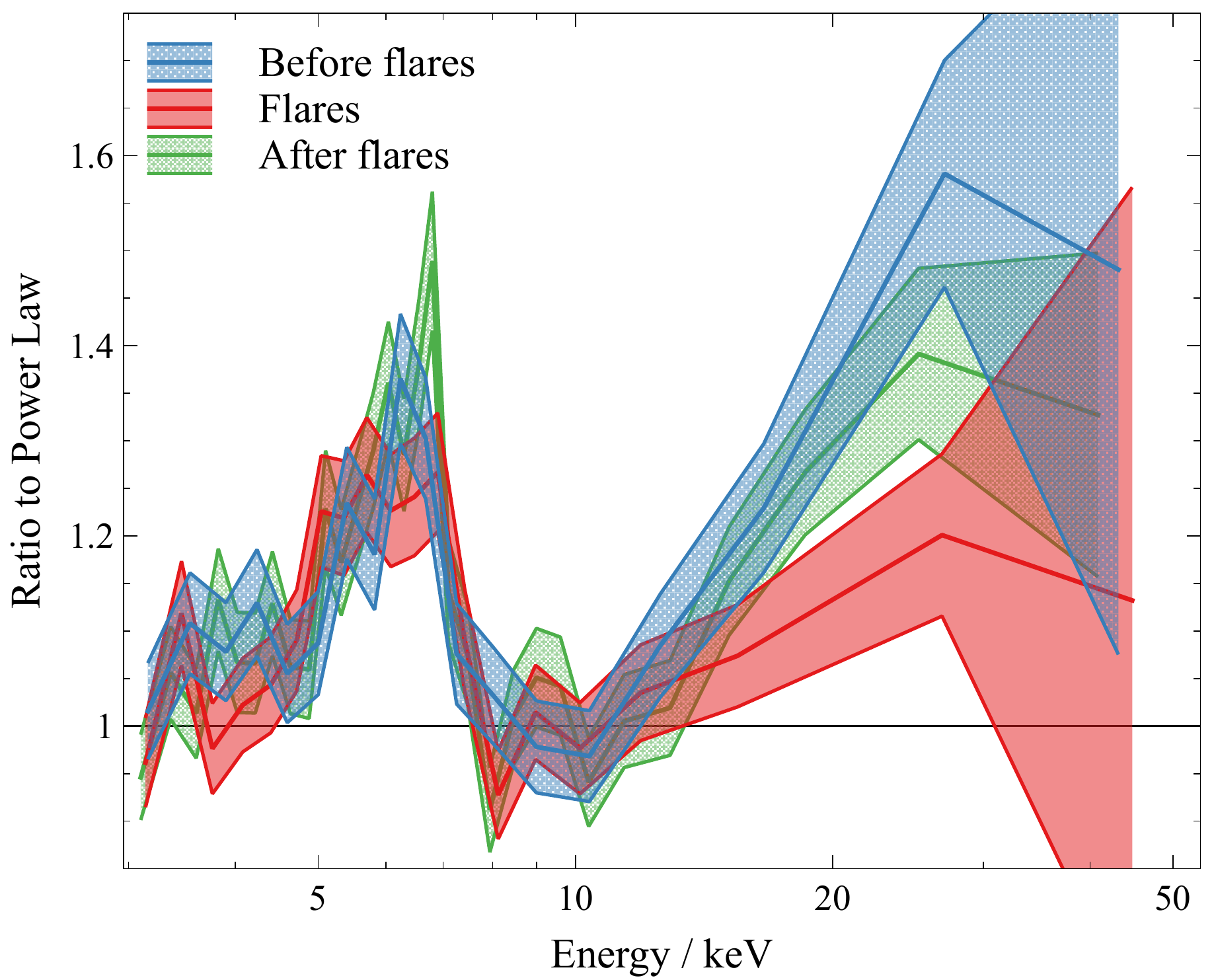}
	\caption{The ratio between the 3-50\keV\ spectrum measured by \textit{NuSTAR} and the best-fitting power continuum, in time intervals before, during and after the flares. The ratio to the power law shows the reflection from the accretion disc: the relativistically broadened iron K line around 6.4\keV\ and the Compton hump. The shape of the reflection spectrum, as well as the reflection fraction (the strength of the reflection relative to the continuum) can be seen to vary.}
	\label{fig:ratio_flare}
\end{figure}

\begin{figure}
	\includegraphics[width=\columnwidth]{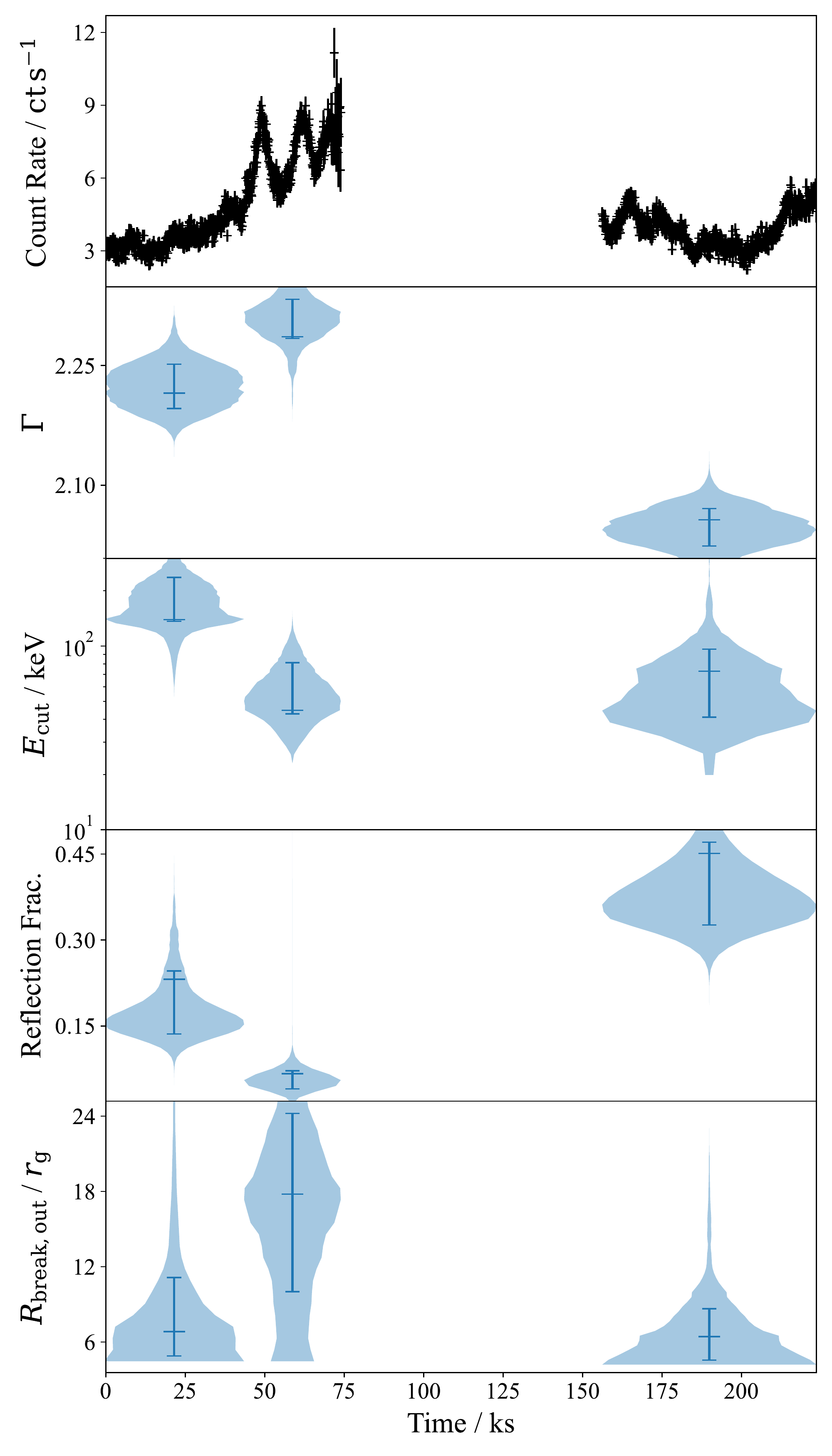}
	\caption{Variation in parameters of the continuum emission and reflection from the accretion disc in time periods before, during and after the flares. The top panel shows the light curves obtained with the \textit{XMM-Newton} EPIC pn camera. The variation is shown in the photon index of the power law continuum spectrum, the continuum cutoff energy, indicating the temperature of the corona, the ratio of the reflected to continuum flux (the reflection fraction), and the outer break radius of the accretion disc emissivity profile, which indicates the extent of the corona over the disc. The width of the bubbles represent the posterior probability density function at each parameter value. The central ticks show the best-fitting parameter values, and the bars represent the $1\sigma$ uncertainties.}
	\label{fig:mcmc_bubbles}
\end{figure}

To understand the underlying changes that lead to the flares observed in the X-ray emission, we compare the best-fitting values of parameters that describe the accretion disc and corona in the time periods before, during and after the flares. The best-fitting parameters to each time period are shown in Table~\ref{tab:param} and the variation in the properties of the corona are summarised in Fig.~\ref{fig:mcmc_bubbles}. Fig~\ref{fig:ratio_flare} shows the changes in the reflection from the accretion disc, illustrated as the ratio between the spectrum observed by \textit{NuSTAR} in the 3-50\keV\ band and the best fitting power law that represents the directly-observed continuum emission. In addition to the tentative evidence that the corona expands during the flares from the change in the profile of the iron K line, we find that the strength of the reflection decreases relative to the continuum during the flares, the X-ray continuum spectrum significantly softens and steepens, and that during the flares the amplitude of the Compton hump decreases with respect to the amplitude of the iron K line.

The weakening of the reflection with respect to the continuum during the flares is manifested in the measured values of the reflection fraction. We find that the reflection fraction drops from $0.23_{-0.11}^{+0.04}$ in the time period before the flares to just $0.07_{-0.04}^{+0.02}$ during the flares, recovering to $0.45_{-0.15}^{+0.03}$ once the flares have subsided. From the posterior distributions of the reflection fraction before, during and after the flares, we can estimate the significance level at which the drop in reflection fraction is detected by testing the null hypothesis that the reflection fraction during the flares is the same or greater than that before or after. We estimate that the null hypothesis that the reflection fraction does not drop as the flares begin can be rejected at the 99.7 per cent ($3.0\sigma$) confidence level, and that the null hypothesis that the reflection fraction does not increase again after the flares can be rejected at the 99.995 per cent ($4.1\sigma$) confidence level.

We see significant variation in the spectrum of the X-ray continuum emitted by the corona softens. The photon index steepens from $2.22_{-0.03}^{+0.04}$ before the flares to $2.29_{-0.02}^{+0.06}$. After the flares subside, the continuum spectrum hardens, with the photon index dropping to $2.06_{-0.04}^{+0.03}$. The increase in photon index during the flares is detected at 99 per cent ($2.6\sigma$) confidence, and the drop in photon index after the flares in detected to at least 99.99999 per cent ($5.4\sigma$) confidence.

In addition to the drop in reflection fraction and the softening of the continuum spectrum, we find tentative evidence that during the flares, the temperature of the corona drops. We find that the cutoff energy of the continuum spectrum drops from $140_{-20}^{+100}$\keV\ before the flares to $45_{-9}^{+40}$\keV\ during the flares, reheating to a cutoff energy of $70_{-30}^{+40}$ after the flares. The drop in the cutoff energy to $\sim 45$\keV\ during the flares can be seen in Fig.~\ref{fig:ratio_flare}. The cutoff energy approximately coincides with the peak of the Compton hump, resulting in the suppression of X-ray emission above this energy and a reduction of the hump relative to the strength of the iron K line during the flares. From the posterior distributions of the cutoff energies before, during and after the flares, we find that the drop in coronal temperature as the flares begin is detected at the 99.96 per cent ($2.7\sigma$) confidence level. As the corona recovers after the flares, the temperature of the corona appears to rise again, but the null hypothesis that the temperature does not increase as the flares end can only be rejected at the 73 per cent ($1.1\sigma$) confidence level.

While the ionisation parameter and power law index of the ionisation profile, along with the column densities and ionisation parameters of the warm absorption components, are allowed to vary between the intervals, we find that the data do not constrain significant variation in these parameters between the three time intervals. We marginalise over these parameters in estimating the uncertainty of each of the parameters of interest.

\subsection{The black hole spin and the inner edge of the disc}

The spin of the black hole can be estimated from the inner radius of the accretion disc assuming that the disc from which the broad iron K line seen extends down to the innermost stable circular orbit (ISCO) in the Kerr spacetime \citep{brenneman_reynolds}. The spin of the black hole was linked between the three time intervals as the spectra were fit simultaneously. The extent of the redshifted wing of the iron K line indicated that the disc extends to the ISCO around a rapidly spinning black hole, with the spin parameter being at least $a>0.94\,GMc^{-2}$ (90 per cent confidence limit). 

As a further test, assuming that the black hole in I\,Zw\,1 is maximally spinning ($a = 0.998\,GMc^{-2}$), we allow the inner radius of the accretion disc to vary freely in each time interval. The posterior distributions of the inner radius of the disc in each of the time intervals allow us to test for truncation of the accretion disc outside the ISCO and to determine whether there is any variation to the inner edge of the disc. We find no evidence for significant variation of the inner edge of the accretion disc. Before the flares, the inner radius of the disc is found to be at a radial coordinate of $3.5_{-0.6}^{+2.6}$\rg. During the flares, the inner radius is most tightly constrained around the ISCO of a rapidly spinning black hole, at $2.1_{-0.4}^{+1.3}$\rg, and is least well constrained after the flares, when we can place an upper limit on the inner radius, lying within 15\rg, though the inner radii during all three time intervals were statistically consistent with one another.

\section{Discussion}
The reflection of X-rays from the inner regions of the accretion disc, measured in the X-ray spectrum of I\,Zw\,1, offers a unique probe of the inner regions of the accretion flow and the corona that produces the X-ray continuum emission. We find that while the 3-50\keV\ X-ray spectrum can be well-described by a simple model consisting of the X-ray continuum and reflection from a disc described by a single ionisation parameter at all radii \citep{1zw1_nature}, in order to describe the broadband 0.3-50\keV\ spectrum, it is necessary to account for the gradient in the ionisation of the disc as a function of radius. The observed reflection spectrum is consistent with a model in which the ionisation parameter falls following a power law in radius. Applying a constant ionisation with variable plasma density suggests that the electron density in the accretion disc may be as high as $\log(n_\mathrm{e}\,/\,\mathrm{cm}^3) = 16.7_{-1.1}^{+0.3}$, but this increased density over the canonical value of $n_\mathrm{e} = 10^{15}$\pcmcu\ is not formally required in the model with a radial ionisation gradient.

Variation of the ionisation parameter with radius is expected when the accretion disc is radiatively ionised by the irradiation from a central, compact corona due to strong variations in ionising flux, as well as the density of the underlying disc, as a function of radius. We find in I\,Zw\,1, however, that accounting for the ionisation gradient is \textit{required} to provide a good description of the spectrum. In the past, phenomenological models consisting of two reflection components have been used to model the broadband spectra of AGN, and such models have been interpreted \textit{post hoc} as accounting for the ionisation structure of the accretion disc. Comparing the DIC statistics, we find that in I\,Zw\,1, the single reflection component with a radial ionisation gradient is preferred to the two-reflector model. Both models produce similar residuals, however the ionisation gradient model has fewer (effective) parameters (and a clearer physical interpretation).

From the profile of the iron K line, we estimate the illumination pattern of the accretion disc by X-rays from the corona \citep[\textit{e.g.}][]{1h0707_emis_paper}. The emissivity profile of the accretion disc is best-described by a twice-broken power law. Before and after the flares, the emissivity profile of the disc is consistent with illumination by a compact corona that extends corona that extends no more than $7_{-2}^{+4}$\rg\ over the inner disc before the flares, and $6_{-2}^{+2}$\rg\ after the flares. During the flares, the emissivity profile is consistent with the corona having expanded to a radius of $18_{-7}^{+6}$\rg\ over the inner parts of the accretion disc, however due to limited exposure time during the flares, this expansion is only tentatively detected and the data are formally consistent with no expansion during the flares. Combining measurements of the emissivity profile with measurements of the reverberation time delay \citep{1zw1_nature}, we may infer that the bulk of the corona extends to a radius of between 6 and 18\rg\ over the accretion disc at an average height of approximately 4\rg\ above the disc plane.

The precise mechanism by which the corona is produced remains unknown. \citet{yuan_fluxtubes_2} show that a compact corona can be formed if the gradient in flux density arising from the disc exceeds a critical value. An instability develops in three-dimensional force-free electrodynamic simulations that causes magnetic field structures that would form a large-scale jet to collapse and dissipate energy in the black hole magnetosphere. This phenomenon can explain compact X-ray sources close to black holes and may be responsible for generating the core of the corona previously inferred from X-ray spectral-timing measurements \citep{1zw1_corona_paper}, or the compact corona seen in I\,Zw\,1 before and after the flares.

\citet{zhu+2020,zhu+2021}, however, find the X-ray emission from most radio-loud AGN likely originates from a corona analogous to that found in radio-quiet AGN, suggesting that there must also be a component of the corona that does not require the destruction of the jet. We may speculate that a component of the corona extending over the innermost radii of the accretion disc (out to around 6\rg\ before and after the flares, but potentially more extended as the flares are launched) can arise from the same magnetic fields that are posited to drive accretion via the disc \citep{balbus+98}. If the field lines responsible for and amplified by the magneto-rotational instability (MRI) accelerate particles and undergo a significant number of reconnection events where the flux is most concentrated, above the innermost parts of the disc, they would naturally be expected to generate X-ray emission.

\subsection{Variation in the coronal temperature}
We find that during the X-ray flares, the temperature of the corona drops, with the high-energy cutoff of the continuum spectrum decreasing from $140_{-20}^{+100}$\keV\  before the flares to $45_{-9}^{+40}$\keV\ during the flares. Although the error bar on the measurement is large during the flares, the decrease is detected at $2.7\sigma$ significance. After the flares, there is tentative evidence that the corona reheats, but increase in cutoff energy is only detected at $1.1\sigma$ significance.

The temperature of the corona determines the high energy cutoff of the power law continuum spectrum and alters the energy balance in the accretion disc as it is heated by the X-ray continuum. We are able to detect the change in coronal temperature because this change in the energy balance within the disc changes the spectrum of the X-rays that are reflected \citep{xillver_temp}. In particular, a lower temperature corona alters the ratio of the brightness of the iron K emission line to that of the Compton hump that are observed in the \textit{NuSTAR} bandpass. A lower energy cutoff reduces the peak of the Compton hump relative to the peak of the line in a way that cannot be explained by variation in the ionisation state of the disc in these data (we marginalise over this parameter in the MCMC calculation). Such a drop in the Compton hump relative to the line can also be due to the iron abundance, however this cannot vary on the timescales upon which flares are observed.

If the coronal X-ray emission is produced by the Comptonisation of seed photons, \textit{e.g.} from the accretion disc \citep{galeev+79}, the drop in the coronal temperature during the flares could be caused either by an increase in the flux of seed photons entering the corona, reducing its temperature by Compton cooling, or by a reduction in the energy density that is deposited by the process that heats the corona. The cooling time of the corona is short, meaning that the continued emission of the X-ray continuum requires constant injection of energy \citep{fabian+2015}. It is likely that the corona is produced by the dissipation of energy from reconnecting magnetic fields associated with the orbiting plasma in the accretion disc and the spinning black hole \citep{merloni_fabian,yuan_fluxtubes_1,yuan_fluxtubes_2,bransgrove+2021}. The characteristic temperature of the corona is therefore likely to be highly sensitive to the same changes to the magnetic field configuration which leads to variation of the location and geometry of the corona. Observing this change in temperature of the corona during the flares gives us new insight into the mechanism by which the corona is formed and by which the flares occur.

In addition to reducing  in the high-energy cutoff, the drop in coronal temperature is consistent with the observed softening of the X-ray continuum. Such behaviour, where the continuum spectrum becomes softer as the X-ray source becomes brighter has previously been observed in AGN \citep{mark_edel_vaughan,1h0707_var_paper,mrk335_corona_paper}, however in these observations of I\,Zw\,1, we are able to directly relate the softening of the continuum spectrum to variations in the temperature of the corona, following previous hints of such behaviour \citep[\textit{e.g.}][]{kang+2021}.

The photon index ($\Gamma$) increased from $2.22_{-0.03}^{+0.04}$ before the flares to $2.29_{-0.02}^{+0.06}$ during the flares.  Following \citet{sunyaev_trumper}, we can estimate the spectrum produced by Comptonisation within a corona with characteristic temperature $T$ and through which the optical depth to Thomson scattering is $\tau$. The spectral index (related to the photon index by $\Gamma = 1 + \alpha$) is given by:
\begin{align}
\alpha &= -\frac{3}{2} + \left( \frac{9}{4} + \gamma \right)^\frac{1}{2},
\end{align}
where
\begin{align*}
\gamma &= -\frac{\pi^2}{A} + \frac{m_\mathrm{e} c^2}{k_\mathrm{B} T\left(\tau + \frac{2}{3}\right)^2}
\end{align*}
The constant $A$ is determined by the geometry of the system. Assuming a spherical geometry, $A=3$. While this relation is formally derived in the diffusion limit, appropriate for high optical depth, it has been shown to be a relatively good approximation in the optically thin limit \citep{titarchuk-94}.

Assuming that the X-ray continuum arises by Comptonisation in a thermal population, and notwithstanding any non-thermal contribution to the X-ray continuum emission during the short-timescale flares, we can use the temperature of the corona measured from the high-energy cutoff to estimate the optical depth that is required through the corona in order to produce a continuum spectrum with the observed photon index. Before the flares (noting the large upper error bar on the temperature measurement), we estimate that $\tau < 0.3$ is required to produce the observed continuum spectrum, consistent with the typical assumption that the X-ray continuum is produced in an optically-thin corona. During the flares, producing the softer continuum spectrum with $\Gamma = 2.29_{-0.02}^{+0.06}$ from a cooler corona with $kT = 45_{-9}^{+40}$\keV\ requires $\tau = 0.8_{-0.4}^{+0.2}$. In the optically-thin limit, the probability of scattering seed photons will scale proportional to the optical depth. This means that, coupled with tentative evidence for the expansion of the corona over the inner accretion disc during the flares, increasing the rate at which seed photons are intercepted, the increase in optical depth by a factor is 2.6 is consistent with the factor 2.5 seen in X-ray count rate during the flares (although a further discussion of the seed photon population can be found below in \S\ref{sec:seed}). After the flares, the hardening of the continuum spectrum to $\Gamma = 2.06_{-0.04}^{+0.03}$ requires $\tau = 0.7_{-0.3}^{+0.4}$. Although the optical depth is greater than that before the flares, contraction of the corona to a smaller area over the disc after the flares reduces the X-ray count rate.

\subsection{The low reflection fraction and motion of the corona}
We find that during the X-ray flares, the reflection fraction drops from $0.23_{-0.11}^{+0.04}$ to $0.07_{-0.04}^{+0.02}$, rising again to $0.45_{-0.15}^{+0.03}$ after the flare. In a simple reflection scenario, in which a compact, isotropically-emitting corona is located above the accretion disc, the reflection fraction is expected to be unity as half of the continuum flux is emitted downwards, toward the disc, and half is emitted upwards to be observed directly. While gravitational light bending focuses rays towards the black hole and inner parts of the disc, increasing the reflection fraction \citep{xmm2015proc}, a reflection fraction below unity indicates that the inner accretion disc is under-illuminated. Such a drop in the reflection fraction can be understood in terms of the corona accelerating away from the accretion disc during the flares. As the corona is ejected away from the disc, relativistic beaming of the X-ray emission means that a greater fraction of the coronal X-rays are emitted upward, away from the disc, hence are observed directly as continuum emission rather than being reflected \citep{beloborodov}. After the flares, the reflection fraction rises again as the remaining corona slows.

A drop in the measured reflection fraction can also be explained by truncation of the inner accretion disc during the flare \citep[\textit{e.g.}][]{demarco+2021}. If the innermost part of the disc is either ejected or becomes optically thin such that no reflection is seen, the solid angle of the reflector subtended at the corona is reduced. Over-ionisation of the inner accretion disc can also cause the disc to appear truncated, if the iron K line emission from the inner disc is weakened and the reflection spectrum is smoothed to the extent that it blends into the continuum. This explanation, however, is inconsistent with the finding that even during the flare, redshifted reflection is detected from radii as far in as $2.1_{-0.4}^{+1.3}$\rg. Ray tracing simulations, following \citet{mrk335_corona_paper}, show that to measure a reflection as low as 0.29 from a static X-ray source at a height of 4\rg, the inner disc would need to be truncated at a radius of 18\rg. The model of the relativistically broadened iron K line, however, probes the inner radius of the accretion disc via the maximal observed redshift of line photons emitted from close to the black hole. Measurements of the inner radius of the accretion disc before, during and after the flare are inconsistent with the disc becoming truncated at larger radius (or over-ionised) during the flare. Indeed, the data quality during the flare are such that the inner radius of the disc is most tightly constrained to be small during the flare.

Assuming that the measured reflection fraction over the 0.3-100\keV\ energy band is representative of total fraction of the 0.3-100\keV\ X-ray continuum that is intercepted by the disc, and that reprocessing of the X-ray continuum emission by the disc does not cause a significant fraction of the incident flux to be re-emitted outside this band, relativistic motion of the X-ray source away from the accretion disc remains the most plausible explanation of such low reflection fractions, and the drop in reflection fraction during the flares. A direct detection of the ejected corona, which would likely be similar in appearance to the ejection of a blob within a jet and be detectable via radio observations, would support this scenario, however no such radio monitoring has yet been conducted simultaneously with X-ray observations of I\,Zw\,1 or of flaring in similar NLS1 AGN. The acceleration of the corona away from the disc during X-ray flares, followed by its subsequent collapse into a confined region around the black hole is consistent with the behaviour previously reported during X-ray flares observed in another NLS1 galaxy, Markarian 335 \citep{mrk335_flare_paper}.

\citet{gonzalez+2017} derive an empirical formula that predicts the reflection fraction as a function of the velocity and height, $z$, of a point-like corona, accounting for not just special relativistic beaming of the emission due to the motion of the corona, but also the effect of light bending close to the black hole focusing emission towards the inner regions of the disc:
\begin{equation}
R(z,\beta) = \frac{\mu'_\mathrm{out} - \mu'_\mathrm{in}}{1 - \mu'_\mathrm{out}}
\end{equation}
Where the emission angles for the limiting rays reaching the inner and outer disc from the source are given by $\mu_\mathrm{out} = 2z / (z^2 + a^2)$ and $\mu_\mathrm{in} = (2 - z^2) / (z^2 + a^2)$, and the prime symbols denote the correction applied to each angle to account for relativistic aberration due to the motion of the source for velocity $\beta = v/c$, with $\mu' = (\mu - \beta) / (1 - \beta\mu)$.

\begin{figure*}
	\includegraphics[width=17.75cm]{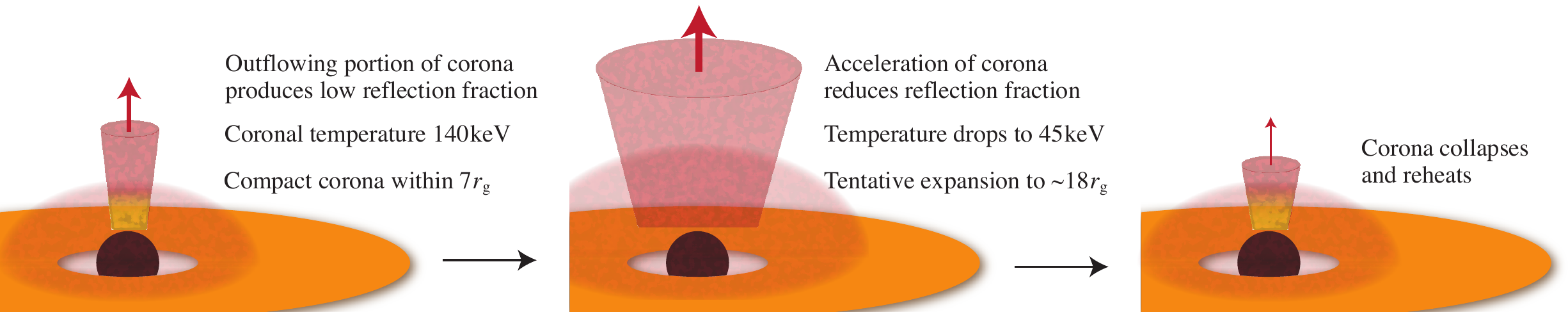}
	\caption{The evolution of the corona during the X-ray flares, inferred from measurements of the reflection fraction, coronal temperature and profile of the broad iron K line over the course of the observations.}
	\label{fig:corona}
\end{figure*}

When an accretion disc around a black hole is illuminated by a compact corona, up to 90 per cent of the reflected flux originates from the innermost 2\rg\ of the disc \citep{1h0707_emis_paper}. A significant fraction of the reflected flux will therefore be lost into the event horizon of the black hole and the observed reflected flux will be reduced to approximately 50 per cent of the total flux emitted form the disc when the X-ray source is between 2 and 10\rg\ from the black hole \citep{return_radiation_paper}. This means that the observed reflection fraction will be approximately half of the `intrinsic' reflection fraction (\textit{i.e.} the ratio of the emitted flux from the corona that reaches the disc to that escaping to be observed directly). Assuming the corona to be a point source at a height of $h=4$\rg\ above the disc plane, as inferred from measurements of the reverberation time delay in I\,Zw\,1 \citep{1zw1_nature}, we can estimate for a reflection fraction of 0.07 to be measured, corresponding to an intrinsic reflection fraction of approximately 0.14, the velocity of the corona during the flare would need to reach $0.89c$, accelerating from a velocity of $0.70c$ just before the flare. The measured reflection fraction after the flare corresponds to a velocity of $0.48c$. It should be noted, however, that these represent upper limits on the velocity, since these estimates assume the corona is a compact point source. Extension of the corona over the inner accretion disc, away from the black hole, will reduce the fraction of rays bent towards the inner disc, thus will reduce the velocity required to produce the observed reflection fraction. The velocity of the corona during the flare exceeds the escape velocity from a radius of $h=4$\rg\ from the black hole, and the plasma ejected from the corona can escape the black hole’s gravitational influence. The evolution of the corona inferred from the variation of the spectrum during the flares is shown in Fig.~\ref{fig:corona}.

If the reflection fraction drops due to the acceleration of the corona away from the disc, it is expected that the apparent redshifting of the disc in the rest frame of the accelerating corona causes the continuum to become photon-starved. This results in a decrease in the photon index as the continuum spectrum hardens \citep{beloborodov}. We, however, detect a \textit{softening} of the continuum spectrum during the flares. This can be explained by the reduction of the coronal temperature we measure during the flares. Decreasing the temperature reduces the Compton amplification factor (the average fractional gain in energy per scattering during the Comptonisation process) and, thus, acts to soften the X-ray continuum.

Based upon the velocities inferred from the reflection fractions, and the relations derived by \cite{beloborodov}, we expect a decrease in photon index by 0.15, corresponding to a factor 10 increase in amplification factor. The amplification factor scales as $(kT)^2$, thus it would only take a factor of three drop in temperature to counteract this hardening of the spectrum, in line with the drop we measure. After the flare, we find that the continuum spectrum hardens. This is consistent with the measured contraction of the corona to a confined region close to the black hole. A compact corona will naturally be photon-starved as much of the solid angle subtended at such a corona is occupied by the event horizon, and relatively little is occupied by the source of seed photons, which is the accretion disc. Such short timescale variability in the temperature of the corona and the photon index of the continuum spectrum it emits likely contributes to the scatter in the observed relationship between the photon index and the Eddington ratio observed in AGN \citep[\textit{e.g.}][]{shemmer+2008,brightman+2013}. While the average photon index can be related to the accretion rate onto the black hole, there is additional short-timescale variability that is not trivially related to the mass accretion rate. The accretion rate likely varies on viscous timescales through the accretion disc, which is much longer than the timescale of the variability we observe.

On the decline of each of the flares, short flashes of photons were detected consistent with the re-emergence of iron K line photons reflected from the back side of the accretion disc, magnified and lensed around the black hole in the strong gravitational field \citep{1zw1_nature}. This interpretation requires that the reflection and reverberation from the accretion disc arises only in response to the first $\sim2000$\s\ of the flare, such that the flashes of photons re-emerging from the back side of the disc remain short. The measured evolution of the spectrum during the flares and the significant drop in the reflection fraction further supports this interpretation. The acceleration of the corona away from the disc and the resultant relativistic beaming of the continuum emission away from the disc means that the inner accretion disc is strongly illuminated only during the first part of the flare when the coronal velocity is low. Once the corona has accelerated at the peak of the flares, the majority of the continuum emission is beamed away from the disc and it no longer produces a strong response.

General relativistic magneto-hydrodynamic (GRMHD) simulations suggest a mechanism for the observed X-ray flares \citet{mckinney+2012} find that, in certain circumstances, large-scale toroidal magnetic flux can be accreted inwards through the accretion disc until a critical flux density is reached in the inner regions. At this point, the inner disc becomes compressed by the magnetic field, which forms a barrier to further accretion, leading to a magnetically choked, or magnetically arrested disc (MAD). Further build-up of magnetic flux can lead to an inversion of the polarity of the field that threads the event horizon, during which magnetic flux is ejected and bursts of rapid accretion occur. The MAD state can also readily arise when geometrically thin discs (as suggested by the appearance of the X-ray reflection spectrum) are highly magnetised \citep{avara+2016}. The ejections of magnetic flux (that would form the corona) and bursts of accretion (with their corresponding radiative output) could plausibly be what we observe as the flares in the X-ray emission. The GRMHD simulations show bursts in mass accretion rate corresponding to the magnetic flux ejections lasting time periods around 400-800$\,GMc^{-3}$ (where $GMc^{-3}$ is the light crossing time across one gravitational radius). In the light curve of I\,Zw\,1 obtained by \textit{NuSTAR}, we see that the flaring lasts 80-100\ks, corresponding to 500-700$\,GMc^{-3}$ (assuming a black hole mass of $3\times 10^7$\Msun, \citealt{vestergaard+06,1zw1_nature}), thus the flaring occurs on approximately the timescale expected in the MAD model.

If the acceleration of the corona is to occur within 2000\s, the process by which it is driven must propagate rapidly through the black hole magnetosphere. 2000\s\ corresponds to a time period of $13\,GMc^{-3}$ for a $3\times 10^7$\Msun\ black hole. From ray tracing calculations around a maximally spinning black hole, $13\,GMc^{-3}$ is the light travel time from a radius of 1.4\rg\ on the accretion disc to a point at height 4\rg\ on the spin axis. If the process driving the flare arises from activity on the inner accretion disc, it must therefore propagate through the magnetosphere at close to the speed of light. In a force-free magnetosphere, in which the kinetic and thermal energy densities of the plasma are negligible compared to the energy density associated with the electromagnetic fields, the Alfv\'{e}n speed, associated with the propagation of waves along magnetic field lines, approaches the speed of light. The rapid nature of the flares points to their origin in the magnetic fields associated with the black hole and inner accretion disc, rather than changes in the accretion disc itself, which would propagate much more slowly on the viscous timescale.

\subsection{The UV light curve and the seed photon population}
\label{sec:seed}
Given that the total X-ray flux is observed to increase during the flares, as the coronal temperature decreases, it is necessary that if the X-ray continuum is produced by Comptonisation, that the flux of seed photons entering the corona increases. This requirement is strengthened if the corona is accelerated away from the disc. Relativistic beaming reduces the solid angle subtended by the accretion disc in the rest frame of the corona, reducing the effective cross section of the corona to photons from the disc, thus additional seed photons must be available to counteract this photon-starving. We can roughly estimate the seed photons available from the accretion disc via the UV light curve, which was recorded by the \textit{XMM-Newton Optical Monitor} (OM) in the `image+fast' timing mode.

\begin{figure}
	\includegraphics[width=\columnwidth]{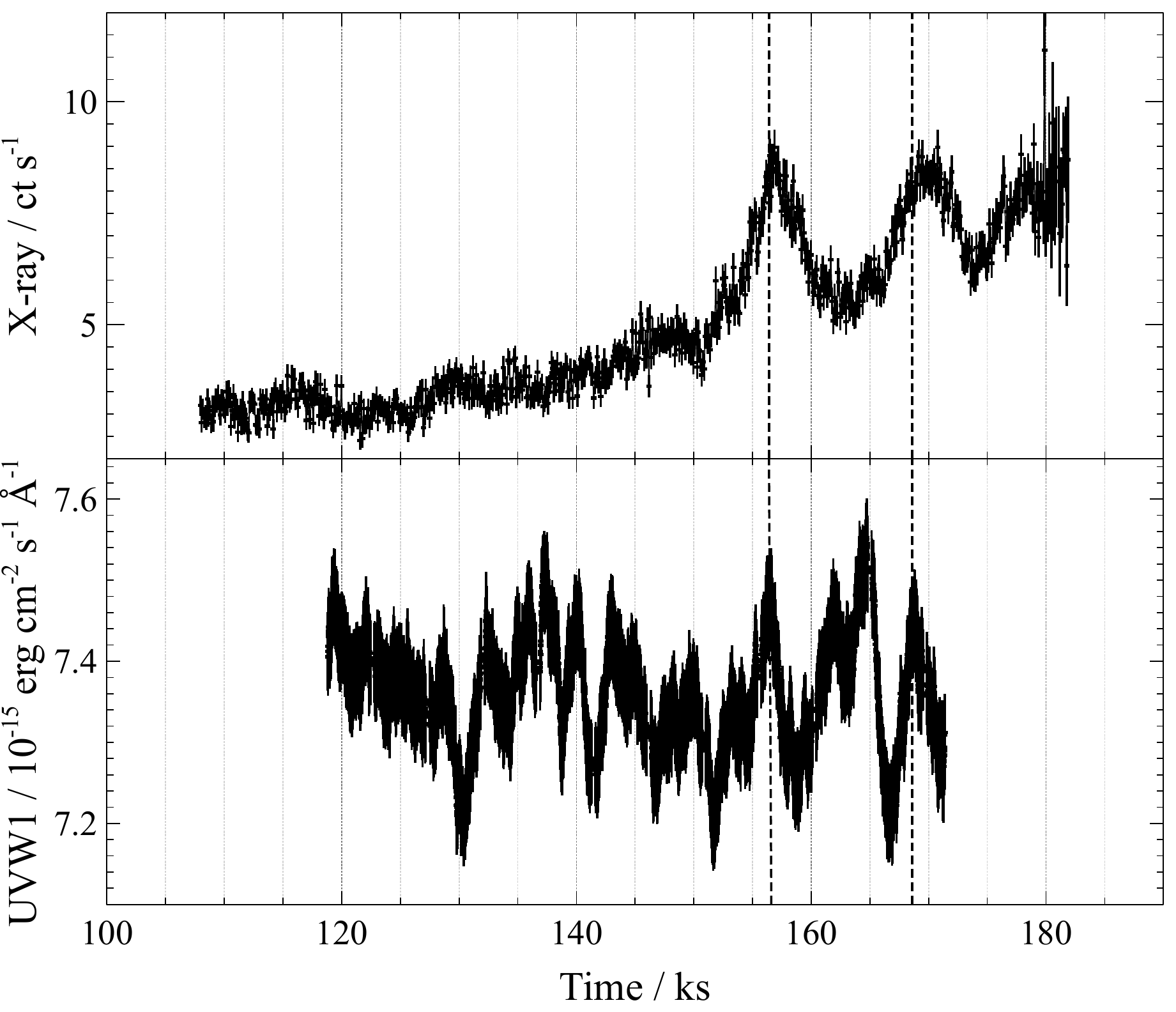}
	\caption{X-ray and UV light curves of I\,Zw\,1, obtained during the first \textit{XMM-Newton} observation using the EPIC pn camera and Optical Monitor (OM), respectively. OM observations were obtained through the \textit{UVW1} filter, with effective wavelength 2910\AA, in `image+fast' timing mode. The raw UV light curve was extracted from the OM data in 10\s\ bins, and a 100-point moving average filter is applied to suppress noise. Vertical dashed lines show peaks in the UV emission that are simultaneous with the peaks of the two X-ray flares.}
	\label{fig:uvlc}
\end{figure}

For the majority of the observations (including the flaring period), OM observations were obtained through the \textit{UVW1} filter, with a central wavelength of 2910\AA. The \textit{UVW1} light curve is shown alongside the X-ray light curve in Fig.~\ref{fig:uvlc}. For a standard accretion disc \citep{shaksun} through which matter is accreting at approximately 20 per cent of the Eddington limit onto a black hole of mass $3\times 10^7$\Msun, emission in the UVW1 band is dominated by regions of the disc around 200\rg\ \citep[see \textit{e.g.}][]{mrk335_flare_paper}, which is likely not the dominant source of seed photons to a corona at smaller radii, within 28\rg\ of the black hole (approximately 1 per cent of the black body emission at 30\rg\ is expected to emerge in the \textit{UVW1} bandpass), so the light curve in this band is only approximate tracer of the available seed photon flux. We note, however, past studies of AGN with the Extreme Ultraviolet Explorer (EUVE) that show that the EUV varies approximately simultaneously with the near-UV, though with up to twice the variability amplitude in the EUV than the near-UV \citep{marshall+1997}.

We see variability at the 4 per cent level in the UV light curve of I\,Zw\,1. Much of the UV variability before the flares shows no obvious correlation with the X-ray flux, but we do note that both of the X-ray flares are simultaneous with peaks in the UV emission, meaning that, in principle, there is a slight enhancement in seed photon flux to produce the X-ray flare through the cooler corona is available, in addition to the tentative evidence for the expansion of the corona such that it possesses a greater cross secion for intercepting seed photons form the disc during the flares. 

It is also of interest that the increase in \textit{UVW1} flux seen peaking 8000\s\ following the peak of the first X-ray flare is consistent with the light travel time from a compact, central corona to a radius of $\sim 150$\rg\ on the disc. It is therefore plausible that this rise in UV flux between the peaks is due to the reprocessing of the flaring X-ray emission heating the surface of the disc. 

\section{Conclusions}
Analysing the broadband X-ray spectrum of the narrow line Seyfert 1 galaxy I\,Zw\,1, measured by \textit{NuSTAR} and \textit{XMM-Newton}, we find that the 0.3-50\keV\ spectrum is well-described by the X-ray continuum emission from the corona, the relativistically-broadened reflection of this X-ray continuum from the accretion disc around the rapidly spinning black hole, and soft X-ray absorption from the two warm, ionised outflows that were previously detected in this AGN.

In order to describe the reflection spectrum over the entire 0.3-50\keV\ energy band, it is necessary to account for the radial variation in ionisation across the accretion disc. Approximating the variation in ionisation parameter as a power law in radius provides a good fit to the observed spectrum.

During a series of X-ray flares, each lasting around 10\ks, the reflection fraction drops, consistent with the acceleration and ejection of the X-ray emitting corona away from the accretion disc. We find the first evidence that the temperature of the corona drops from $140_{-20}^{+100}$\keV\ before the flares to $45_{-9}^{+40}$\keV\ during the flares. This drop in the temperature of the corona is consistent with the observed softening of the X-ray continuum during the flares, and subsequent hardening after the flares subside.  In a model in which the X-ray continuum is produced by the Comptonisation of thermal seed photons from the accretion disc, we infer changes in the optical depth by a factor of approximately 2.5

Before and after the flares, the emissivity profile of the accretion disc, measured via the profile of the broad iron K line, and the reverberation time lag between the continuum and the iron K line, are consistent with the inner accretion disc being illuminated by a compact corona, extending no more than $7_{-2}^{+4}$\rg\ over the disc before the flares, and $6_{-2}^{+2}$\rg\ over the disc after the flares. There is tentative evidence that during the flares the corona extends to $18_{-7}^{+6}$ over the inner accretion disc, as it is accelerated away, increasing the cross section for scattering seed photons from the disc. The combination of the expansion and cooling of the corona, the increased optical depth and the peaks in the UV seed photon flux seen during the flares is consistent with the production of the flaring X-ray continuum by the Comptonisation of photons from the accretion disc as they pass through the corona.

From the reflection and reverberation of X-rays off of the inner regions of the accretion disc, a picture is starting to emerge of the extreme environment around the innermost stable orbit and just outside the event horizon of the black hole. The reflection spectrum and reverberation time delays between variations in the continuum and in the emission lines produced from the accretion disc reveal the properties of the X-ray source, including its location and extent. By observing the changes to the corona that underlie extreme episodes of variability, namely the bright X-ray flares emitted by I\,Zw\,1, we can begin to understand the physical processes underlying its formation. Observing how the corona accelerates and cools during the flares and then collapses the flares subsides provides new constraints on the mechanism by which the corona is formed and energised by the accretion flow and black hole, and by which accretion onto supermassive black holes is able to power some of the most luminous objects in the Universe. The rapid timescale of the flares points to their origin in the magnetic fields associated with the accretion disc and black hole.

\section*{Acknowledgements}
This work was supported by the NASA \textit{NuSTAR} and \textit{XMM-Newton} Guest Observer programs under grants 80NSSC20K0041 and 80NSSC20K0838. DRW received additional support from a Kavli Fellowship at Stanford University. WNB acknowledges support from the V.M. Willaman Endowment. Computing for this project was performed on the Sherlock cluster. DRW thanks Stanford University and the Stanford Research Computing Center for providing computational resources and support. We thank the anonymous referee for their valuable feedback on the original version of this manuscript.


\section*{Data Availability}
The data used in this study are available in the \textit{NuSTAR} and \textit{XMM-Newton} public archives. The \textit{NuSTAR} observations can be accessed through the NASA HEASARC (\url{https://heasarc.gsfc.nasa.gov}) via observation ID 60501030002. The \textit{XMM-Newton} observations can be accessed via the XMM Science Archive (\url{http://nxsa.esac.esa.int/nxsa-web}) via observation IDs 0851990101 and 0851990201. Analysis of the X-ray spectra was conducted using \textsc{xspec}, distributed as part of the \textsc{heasoft} package (\url{https://heasarc.gsfc.nasa.gov/docs/software/heasoft}). The \textsc{relxill} and \textsc{xillver} X-ray reflection models are available at \url{http://www.sternwarte.uni-erlangen.de/~dauser/research/relxill}. The \textsc{kdblur3} relativistic blurring kernel, using a twice-broken power law emissivity profile, is available at \url{https://github.com/wilkinsdr/kdblur3}. Scripts to conduct more detailed aspects of the analysis presented here, including the measurement of the accretion disc emissivity profile, are available upon request to the corresponding author.

\bibliographystyle{mnras}
\bibliography{agn}


\bsp	
\label{lastpage}
\end{document}